\documentclass[aps,prc,twocolumn,showpacs,superscriptaddress,floatfix]{revtex4-1}

\usepackage{dcolumn}
\usepackage{amsmath,amssymb,amsfonts}

\usepackage{graphicx}

\usepackage[section]{placeins}
\usepackage{float}
\usepackage{url}
\usepackage[T1]{fontenc}
\usepackage{times} 
\usepackage{morefloats} 
\usepackage[margin=5pt, font=footnotesize, justification=raggedright, format=hang ]{caption}

\begin{document}


\title{Studies of the Three-Body Breakup in Deuteron-Deuteron Collisions \\near the Quasi-Free Limit at 160 MeV}

\author{I.~Ciepa\l}
\email[Electronic mail: ]{izabela.ciepal@ifj.edu.pl}
\affiliation{Institute of Nuclear Physics, PAS, PL-31342 Krak\'ow, Poland}
\author{G.~Khatri}
\affiliation{Department of Physics and Astronomy, Northwestern University, Evanston, IL 60208, USA}
\author{K.~Bodek}
\affiliation{Institute of Physics, Jagiellonian University, PL-30348 Krak\'ow, Poland}
\author{A.~Deltuva}
\affiliation{Institute of Theoretical Physics and Astronomy, Vilnius University, Vilnius, Lithuania}
\author{A. C.~Fonseca}
\affiliation{Centro de F\'{\i}sica Nuclear da Universidade de Lisboa, P-1649-003 Lisboa, Portugal}
\author{N.~Kalantar-Nayestanaki}
\affiliation{KVI-CART, University of Groningen, NL-9747 AA Groningen, The Netherlands}
\author{St.~Kistryn}
\affiliation{Institute of Physics, Jagiellonian University, PL-30348 Krak\'ow, Poland}
\author{B.~K\l os}
\affiliation{Institute of Physics, University of Silesia, PL-41500 Chorz\'{o}w, Poland}
\author{A.~Kozela}
\affiliation{Institute of Nuclear Physics, PAS, PL-31342 Krak\'ow, Poland} 
\author{J.~Kubo\'{s}}
\affiliation{Institute of Nuclear Physics, PAS, PL-31342 Krak\'ow, Poland}    
\author{P.~Kulessa}
\affiliation{Institute of Nuclear Physics, PAS, PL-31342 Krak\'ow, Poland}               
\author{A.~\L{}obejko}
\affiliation{Institute of Physics, University of Silesia, PL-41500 Chorz\'{o}w, Poland}
\author{A.~Magiera}
\affiliation{Institute of Physics, Jagiellonian University, PL-30348 Krak\'ow, Poland}
\author{J.~Messchendorp}
\affiliation{KVI-CART, University of Groningen, NL-9747 AA Groningen, The Netherlands}
\author{I.~Mazumdar}
\affiliation{Tata Institute of Fundamental Research, Mumbai 400 005, India}
\author{W.~Parol}
\affiliation{Institute of Nuclear Physics, PAS, PL-31342 Krak\'ow, Poland}
\author{D. Rozp\k{e}dzik}
\affiliation{Institute of Physics, Jagiellonian University, PL-30348 Krak\'ow, Poland}
\author{I.~Skwira-Chalot}
\affiliation{Faculty of Physics University of Warsaw, PL-02093 Warsaw, Poland}                 
\author{E.~Stephan}
\affiliation{Institute of Physics, University of Silesia, PL-41500 Chorz\'{o}w, Poland}
\author{A. Wilczek}
\affiliation{Institute of Physics, University of Silesia, PL-41500 Chorz\'{o}w, Poland}  
\author{B.~W\l{}och}
\affiliation{Institute of Nuclear Physics, PAS, PL-31342 Krak\'ow, Poland}    
\author{A. Wro\'{n}ska}
\affiliation{Institute of Physics, Jagiellonian University, PL-30348 Krak\'ow, Poland}
\author{J.~Zejma}
\affiliation{Institute of Physics, Jagiellonian University, PL-30348 Krak\'ow, Poland}

\date{\today}

\begin{abstract}
A set of differential cross section of the three-body $^{2}$H($d$,$dp$)$n$ breakup reaction at 160 MeV deuteron 
beam energy are presented for 147 kinematically complete configurations near the quasi-free scattering kinematics.
The experiment was performed at KVI in Groningen, the Netherlands  
using the BINA detector. The cross-section data have been normalized to the $^{2}$H($d$,$d$)$^{2}$H elastic scattering 
cross section. The data are compared to 
the recent single-scattering approximation (SSA) calculations for three-cluster breakup
in deuteron-deuteron collisions. 
Confronting the SSA predictions with the experimental data shows that  SSA provides 
the correct order of magnitude of the cross-section data.
The studied energy is probably too low to meet the SSA assumptions which prevents better accuracy of the description.
\end{abstract}


\maketitle

\section{Introduction}
\label{intro}
After a long experimental campaign searching for three-nucleon force (3NF) effects 
in three-nucleon systems (3N) at intermediate energies, the attention is nowadays directed to heavier systems composed of four nucleons (4N).\\
\indent In the last decades, various final states of {\em N-d} and {\em d-N} scattering were under extensive investigations delivering 
high-precision data for elastic scattering, breakup and radiative capture reactions
for large energy and phase-space ranges \cite{Sag:10,Kal:12,Kis:13}. Together with rigorous Faddeev calculations
for the 3N system the data constitute a sensitive tool to study dynamics of nuclear systems.
Among all the reactions studied, the breakup leading to a final state with three free particles,  
offers the richest phase space with continuum of the final states and is the leading channel at intermediate energies.
A large amount of  possible kinematic configurations makes possible a systematic study of various 
dynamical effects like 3NF, Coulomb force between protons, or relativistic effects, which manifest themselves locally with 
different strength. These features make the breakup reaction a very sensitive and simultaneously strict 
tool for validation of the theoretical models.\\
\indent The present-day models of nucleon-nucleon (NN) forces are based on the meson-exchange theory,
which stems from Yukawa's idea \cite{Yuk:35}. The new generation NN potentials like 
Argonne V18 (AV18) \cite{Wir:95}, CD Bonn (CDB) \cite{Mach:01}, Nijmegen I and II  \cite{Stoks:94} reproduce 
the NN data with extremely high precision,
expressed by $\chi^{2}$ per degree of freedom very close to one. These so-called realistic
NN forces are used in 3N Faddeev equations \cite{Faddeev:61} together with current models of 3NF like  
Urbana IX \cite{Pudli:97},  Tucson-Melbourne (TM99) \cite{Coon:01} and a coupled-channel 
potential CD Bonn+$\Delta$ \cite{Del:03,Del:08} with a 
$\Delta$-isobar degree of freedom, delivering exact solution of the 3N scattering problem. 
These studies are complemented by calculations based on chiral perturbation theory (ChPT)\cite{Ent:03,Ent:17}, 
which are expected to provide
in future consistent description of 2N, 3N (4N etc.) forces.\\
\indent In general the NN potentials supplemented with the 3NF models give a better 
agreement between the proton-deuteron cross-section data and the calculations \cite{Kis:03,Kis:05,Kis:06,Cie:15},
whereas many problems are found for the spin observables \cite{Sekiguchi:09,Stephan:10}. 
This yields a conclusion that the spin part of 3NF is still not under control in the theoretical models. 
It seems that at intermediate and higher energies the inclusion of 3N forces 
is necessary \cite{Kis:03,Kis:05,Cie:15,Esl:09,Sekiguchi:09,Stephan:10,Kal:12,Kis:13} 
but available models are not sufficient.
At lower energies persistent discrepancies exist such as the A$_{y}$ puzzle or the space-star anomaly \cite{Sag:10}. \\
\indent The 4N systems constitute another large and important basis for studies of the 3N forces.
Naively, one can expect the 3NF effects to be increased in the 4N system 
due to the fact that the number of 3N combinations with respect to
2N combinations gets larger with rising the number of nucleons. However, due to expected short
range of 3NF, for large nuclei, the saturation of 3NF effects sets in very quickly. 
4N systems are very suitable to study the 3NF dependence on spin and isospin in the low-energy regime
due to existence of numerous resonance states of different spin and isospin structure. 
The 4N systems are also more flexible to test various nuclear potentials 
in an isospin-dependent way \cite{Ester:13,Hiy:16}.\\
\indent This makes the experimental studies attractive, however the theoretical 
treatment of 4N scattering at medium energies (well above the breakup threshold) 
is much more complicated and challenging than for 3N systems.   
The developments on the 4N field are mainly due to the work of three groups: Pisa \cite{Viviani:11, Kiev:08},
Grenoble-Strasbourg \cite{Lazauskas:04, Lazauskas:09}, and Lisbon-Vilnius \cite{Del:08,Del:14a,Del:15b}. 
Only the Lisbon-Vilnius group 
calculates observables for multi-channel reactions above the breakup threshold, and with the Coulomb
force included. They use the momentum space equations of Alt, Grassberger and Sandhas (AGS)  for transition operators 
in contrary to the two other groups which use the coordinate-space representation.\\
\indent Recently, the calculations were extended 
for higher energies, above the four-cluster breakup threshold, up to an energy of 35~MeV. 
The following models were utilized  in the calculations:
CD Bonn (CDB) \cite{Mach:01} and Argonne V18 (AV18) \cite{Wir:95} potentials, 
INOY04 (the inside-nonlocal outside-Yukawa) potential by Doleschall \cite{Dol:04}, potential
derived from ChPT at next-to-next-to-next-to-leading order (N3LO) \cite{Ent:03},
the two-baryon coupled-channel potential CD Bonn+$\Delta$ \cite{Del:03}. The last model yields effective 
three- and four-nucleon forces \cite{Del:08}, but their effect is of moderate size at most. 
The sensitivity to the force model in the energy range studied reached 30\% in the cross section minimum.
The predictions have been made for observables in $p$-$\!^{3}\mathrm{He}$ \cite{Del:13}, $p$-$\!^{3}\mathrm{H}$, $n$-$\!^{3}\mathrm{He}$ \cite{Del:14a} 
elastic and transfer reactions. 
Recent progress in calculations for {\em d}+{\em d} system is presented in Refs. \cite{Del:15b, Del:16, Del:17}.\\
\indent  The first estimate calculations for the {\em d}+{\em d} system at higher energies 
are currently feasible and were performed in the so-called single-scattering 
approximation (SSA) for the three-cluster breakup 
and elastic scattering \cite{Del:16}. 
In this approximation instead of solving the full AGS equations~\cite{Gras:67}
the 4N operators are expanded in Neumann series in terms of 3N transition operators and 
only the first-order contribution is retained. 
This simplification could be expected to give reasonable predictions only near quasi-free scattering (QFS) kinematics
and with high enough relative {\em n}-{\em d} and {\em n}-{\em p} energies what implies relatively high beam energies.\\
\indent Two types of calculations were performed \cite{Del:16}.
The first one, the so-called one-term ({\em 1-term}) SSA, refers to a situation 
in which the target deuteron breaks due to its proton interaction with the deuteron beam.
In this case the differential cross section is peaked at the neutron spectator energy E$_{n}$=0.
The second one, the so-called four term ({\em 4-term}) SSA, on top of the {\em 1-term} SSA contains 
other three contributions, one of them corresponding to 
the case in which not the target proton but the neutron interacts with the beam deuteron. Two further contributions
arise exchanging the target and beam deuterons, i.e., they correspond to the breakup of the beam deuteron. 
Since the  Coulomb force and interaction in two out of three pairs of three final clusters {\em d}, {\em p}, and {\em n} is neglected,
the  relative energy between those clusters should be high enough to reduce the effects of the final state interaction.
An agreement between {\em 1-term} and {\em 4-term} calculations indicates that the {\em 1-term} reaction mechanism 
dominates in the scattering. The disagreement is a hint of a more complicated reaction mechanism and behavior beyond SSA \cite{Del:16}.\\
\indent To investigate the reliability of the SSA calculations, 
a similar approximation was applied to {\em p}+{\em d} breakup.
The SSA calculations were compared with the exact ones \cite{Del:16}. 
The total {\em p}+{\em d} breakup cross section calculated in an exact way is lower than the one obtained in 
SSA by 30\% at 95 MeV and by 20\% at 200 MeV. The authors of \cite{Del:16} conclude that the SSA should provide 
correct orders of magnitude for total and differential cross sections for {\em d}+{\em d} and {\em p}+{\em d} breakup (near quasi-free region) 
and elastic scattering. \\
\indent Since $n$-$\!^{3}\mathrm{He}$ experiments are difficult, the $p$-$\!^{3}\mathrm{He}$ and  {\em d}+{\em d} reactions
dominate in measurement for the 4N system. The theoretical calculations for the $p$-$\!^{3}\mathrm{He}$ system are the simplest, 
since only elastic and breakup channels exist. 
On the other hand, the most serious complication arise from the Coulomb 
interaction between protons, which is treated using the method of screening and renormalization \cite{Del:13}.
The database for the 4N systems consist of few measurements 
for the elastic  \cite{Alde:78,Mich:07,Bail:09,Ram:11}, breakup \cite{Val:72,Klu:78,Cow:74,Yua:77,Wiel:81,Mye:83} 
and transfer channels \cite{Bizard:80}. In the breakup sector the existing data 
are usually limited to low energies and only very few selected configurations. The new-generation 
data covering large phase space were measured at KVI at 130 \cite{Ram:11,Ram:13} and 160 MeV (this paper). The data evaluation was focused 
on QFS, with neutron acting as a spectator. 
The breakup analysing power data for the $^{2}$H({\em d},{\em dp}){\em n} at 130 MeV were compared with the elastic {\em d}-{\em p}
scattering \cite{Ram:11,Ram:13}.
Recently, the data have been also compared to the SSA calculations and large discrepancy of a factor 1000 was observed 
for differential cross section~\cite{Del:16}, indicating a need to revise both theory and data.\\
\indent In this paper a rich set of the $^{2}$H({\em d},{\em dp}){\em n} differential cross section near the QFS region at 160 MeV deuteron 
beam energy is presented. The data are compared with the SSA calculations \cite{Del:16}.

\section{Detector and experimental technique}
\label{exp}
The experiment was carried out at Kernfysisch Versneller Instituut
(KVI) in Groningen, the Netherlands. The deuteron beam was provided by the superconducting
cyclotron AGOR (Accelerator Groningen ORsay) at kinetic energy of 160
MeV and was impinging on a liquid deuterium target with the nominal thickness of 6.0 mm.
Low beam current (about 5 pA) was used in order to keep the level of accidental
coincidences as low as possible. The reaction products were detected using Big Instrument
for Nuclear Polarization Analysis (BINA) \cite{Mar:08,Stephan:13} designed to study few-body scattering reactions at medium energies.
The BINA setup allows to register coincidences of two-charged particles in nearly 4$\pi$ solid angle, making possible
studies of breakup and elastic scattering reactions. The detector is
divided into two main parts, the forward Wall and the backward Ball.
\begin{figure}[!th]
  \includegraphics[width=0.4\textwidth]{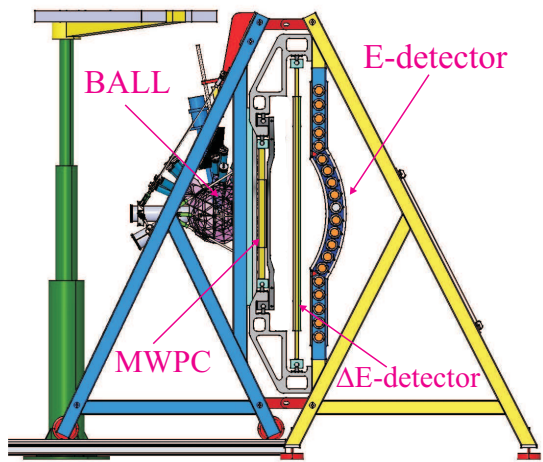}
 \caption{A schematic view of the BINA detector.}
 \label{fig1}
 \end{figure}
A schematic view of the detection system is presented in Fig. \ref{fig1}.
\subsection{Forward Wall}
The forward Wall is composed of a three-plane multi-wire proportional chamber (MWPC)
and an array of an almost-square-shaped $\Delta E$-{\em E}  telescopes formed by 
two crossed layers of scintillator hodoscopes (vertically placed
thin transmission-$\Delta E$ strips and horizontally placed thick stopping-{\em E} bars). The forward
Wall covers polar angles $\theta$ in the range of 10$^\circ$-35$^\circ$ with the full range of azimuthal angles $\varphi$. 
MWPC is used to determine the position of the passing particle. Taking into account the target and beam sizes, the 
accuracy of the angle reconstruction is 0.3$^{\circ}$ for $\theta$ and between 0.6$^{\circ}$ and 3$^{\circ}$ for $\varphi$. 
$\Delta E$ and {\em E} detectors are
used for measuring the energies of the charged reaction products and facilitate the particle identification.  
The energy resolution is about 2\%. MWPC and the hodoscopes, have a central hole to allow for the passage of
beam particles to the beam dump. \\
\indent For BINA the electronic, read-out and data acquisition systems were adopted from its predecessor, 
the SALAD detector \cite{Stephan:09}. The data were collected with various trigger 
conditions to selectively enhance coincidences from the studied reaction channels. 
The trigger conditions were based on hit multiplicities in left side photomultipliers (PM's)
of Wall, right side PM's of Wall and PM's of Ball. Three type of events were registered:
Wall-Wall coincidences, Wall-Ball coincidences and single-type events with at least one particle detected 
in the whole setup. The single-type events were strongly downscaled. \\
\indent The results presented in this paper were obtained only on the basis of the data registered 
in the Wall part of the detector.

\section{Data analysis}
\subsection{Reconstruction of particle trajectories}
\label{data1}
The tracks were built for each event, starting
from hits in the MWPC wire-planes. Their correspondence with the $\Delta E$ and {\em E} detectors was checked
in a track reconstruction procedure. Only events with the consistent information in all three detectors
forming the track are likely to represent charged particles. 
In the analysis two kinds of tracks were considered to estimate the systematic errors connected to the reconstruction: 
the so-called complete and weak tracks with the three and two responding MWPC planes, respectively.  
For complete tracks the reconstruction of angles has been improved, in comparison 
to the previous approaches \cite{Kis:03,Kis:05,Stephan:10}, 
by taking into account also an active wire in the {\em U}~-~plane. 
Consequently, the position resolution was improved by a factor of 1.5.\\
In the case of the weak tracks based on information from {\em U}~-~plane, 
the position resolution in either horizontal or vertical direction is worsen by a factor of 1.2 
as compared to the tracks reconstructed on the basis of {\em X} and {\em Y}-planes. 
Two classes of events were accepted for further analysis:
single-track and two-track events. Tracks with missing
MWPC or $\Delta E$ hits, but with hit in {\em E}, were used to calculate
detector efficiencies. Knowing the crossing point of the responding MWPC wires and
distances between the target and the wire-planes and assuming particle emission from
the point-like target, the polar ($\theta$) and azimuthal ($\varphi$) scattering
angles in the laboratory frame were reconstructed.\\
\indent To calculate configurational efficiency of the {\em E} or $\Delta E$ detectors for coincident events, 
discussed further in Sec.~\ref{effic2}, so-called {\em particular tracks} were reconstructed. 
Exactly two sets of {\em X}-{\em Y}-{\em U} hits in the three MWPC planes matching 
with a single {\em E}-bar (or a single $\Delta E$-strip)
were required. 

\subsection{Particle identification and energy calibration}
\label{data2}
\begin{figure}[!h]
 \includegraphics[width=0.4\textwidth]{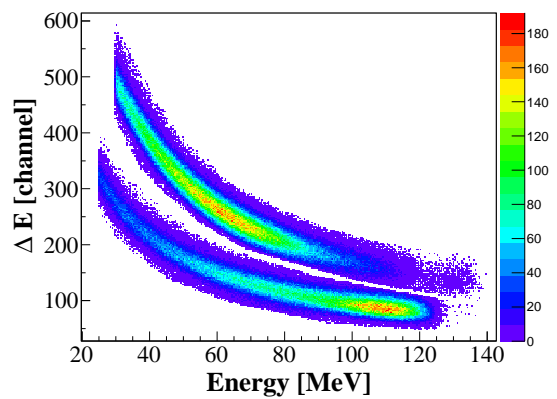}
 \caption{(Color online) Example of the $\Delta E$-{\em E} identification spectrum
 drawn for proton-deuteron coincidences built under graphical cuts conditions. 
 Protons and deuterons branches are well contained within the applied cuts.
 The sharp edges in the both branches, at the lowest energies, are due to a detection threshold.}
 \label{fig2}
 \end{figure}

In order to select the events of interest, proton-deuteron pairs from the breakup reaction,
the $\Delta E$-{\em E} particle identification (PID) technique was applied for each individual 
telescope. The protons and deuterons 
branches were selected by graphical cuts which define an arbitrary area ("banana" shape), wide enough to avoid significant
losses of events. 
A sample identification spectrum built of two-track events is presented in Fig.~\ref{fig2}.\\
\indent After introducing PID into the analysis, the energy calibration was performed 
for each type of particles. In the case of the Wall detector only the stopping {\em E}-detector was calibrated.
Each {\em E}-bar is equipped with two photomultiplier tubes (PMTs) on
its two ends (left-PMT and right-PMT). When a charged particle hits the {\em E}-detector,
in an ideal situation both left and right PMTs respond, giving two pulse height values (C$_{L}$, C$_{R}$). The gains of PMT's 
were well matched, so the difference between C$_{L}$ and C$_{R}$
is mainly due to different light attenuation along the path in the scintillator. 
To perform the energy calibration for a given {\em E}-bar, a nearly position independent C$_{LR}$ variable was obtained
as a geometric mean of C$_{L}$ and C$_{R}$, i.e.
C$_{LR}$ = $\sqrt{C_{L}\times C_{R}}$. The exponential attenuation 
component cancels in the C$_{LR}$ value. The central two {\em E}-detectors were partially cut in the middle to accommodate beam line 
and in this case a sum of the two signals, i.e. C$_{LR}$ = C$_{L}$+C$_{R}$ was used. 
To calibrate the detectors the positions of the peaks corresponding to 
protons originating from the {\em d}-{\em p} elastic scattering process measured in dedicated runs with the 
energy degraders were used  \cite{Parol:14a}. They were compared with the results
of simulations taking into account energy losses of particles along their trajectory. 
In the first step a non-linear function was fitted to the relation of the
deposited energy versus pulse-height. Such a non-linear character (below 40 MeV)
is caused by the quenching effect in the scintillating material
(described with the Birk's formula \cite{Bir:51}). The calibration functions were obtained 
at each polar angle, for left- and right-half sides of a given {\em E}-detector.
In the next step of the calibration the relation between the energy deposited
by protons (or deuterons) in the {\em E}-detector and their energy at the reaction point was 
found with the use of the dedicated GEANT4 simulations of the energy losses in the BINA setup.
Due to different scintillation light output for protons and deuterons,
additional corrections were introduced to the deuteron calibration which were based 
on a well known light output to energy deposit relations. 

\subsection{Detector efficiency}
\label{effic}
To calculate the absolute values of the cross section, it is necessary to take into account the inefficiency of the detectors.
In the case of the BINA setup, the largest inefficiency was related to the detection of particles in MWPC. 
During the experiment certain channels were malfunctioning or ceased to function at all (``dead'' wires) 
decreasing the overall efficiency. Dependence of the MWPC efficiency on the energy deposition 
($E_{loss}$) of a particle in this detector was observed and taken into account. \\
\subsubsection{MWPC efficiency}
\label{effic1}
\begin{figure}[!h]
 \includegraphics[width=0.45\textwidth]{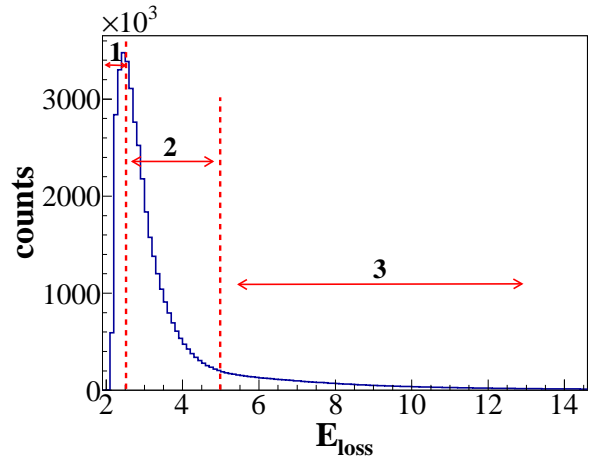}
 \caption{(Color online) The energy loss $E_{loss}$ distribution for the charged particles registered in the BINA setup. 
 The three ranges in the $E_{loss}$ variable in which the MWPC efficiency maps were calculated are depicted.}
 \label{fig2a}
 \end{figure}

\indent In order to correct the numbers of registered proton-deuteron coincidences from the breakup reaction
 and elastically scattered deuterons,
energy loss dependent efficiency maps were constructed, see also \cite{Cie:18}.
The detector acceptance was divided into bins in azimuthal and 
polar angles and efficiency was calculated for each cell with 
the use of the registered single particle events (protons and deuterons). 
Protons and deuterons were treated together to increase the precision of efficiency, and their energy loss $E_{loss}$
were recalculated per unit distance according to the formula:
\begin{eqnarray}
E_{loss}=Q^{2}\cdot\frac{\alpha \cdot m}{T}
\label{eq1}
\end{eqnarray}
where {\em Q} is proton or deuteron charge, {\em m} is the mass of the particle, {\em T} is its kinetic energy
and $\alpha$ is an arbitrary constant factor. The efficiency maps were calculated for three 
ranges in the $E_{loss}$ variable shown in Fig. \ref{fig2a}.
The active part of MWPC contains three planes: {\em X}, {\em Y} and {\em U}, respectively with vertical, 
horizontal and inclined by 45$^{\circ}$ wires. 
The position-sensitive efficiency of each plane was obtained using 
the information from the remaining two others \cite{Parol:14, Cie:18}
and combined with the information from the scintillator hodoscopes.
The probability of registering a particle in a given MWPC plane, e.g. in the {\em X} plane, 
for a given angular bin ($\theta$, $\varphi$) and $E_{loss}$ range is given as:
\begin{eqnarray}
\varepsilon_{x}(\theta,\varphi,E_{loss})=\frac{N_{xyu}(\theta,\varphi,E_{loss})}{N_{yu}(\theta,\varphi,E_{loss})}
\label{eq1}
\end{eqnarray}
where $N_{xyu}(\theta,\varphi,E_{loss})$ is the number of tracks registered in this angular bin 
with at least one wire hit in each of {\em X, Y}, and {\em U} planes, whereas  
$N_{yu}(\theta,\varphi,E_{loss})$ is the number of tracks with at least one wire hit in plane {\em Y}
and one in plane {\em U}. The efficiencies of {\em Y} and {\em U} planes were calculated in a
similar way. The overall MWPC efficiency was obtained as a product 
of the particle registration probabilities in the individual planes:
\begin{eqnarray}
\varepsilon_{xyu}&=&\varepsilon_{x}\cdot\varepsilon_{y}\cdot\varepsilon_{u}.
\label{eq2}
\end{eqnarray}
The MWPC efficiency $\theta$ vs. $\varphi$ maps for the three $E_{loss}$ ranges are presented in Fig.~\ref{fig3abc}. The efficiency is the highest
for the slowest particles which lose more energy in the detector so the signal is well above the applied thresholds.
The difference in the total efficiency between the two maps is about 10\%.\\
 Similar map was obtained also for the so-called weak tracks which allowed for one plane without hit.  
  They were found to be much less sensitive to the energy loss of the particles.
   In this case the MWPC efficiency was calculated for the whole range of $E_{loss}$ according to the formula:
  \begin{figure}[!bh]
 \includegraphics[width=0.38\textwidth]{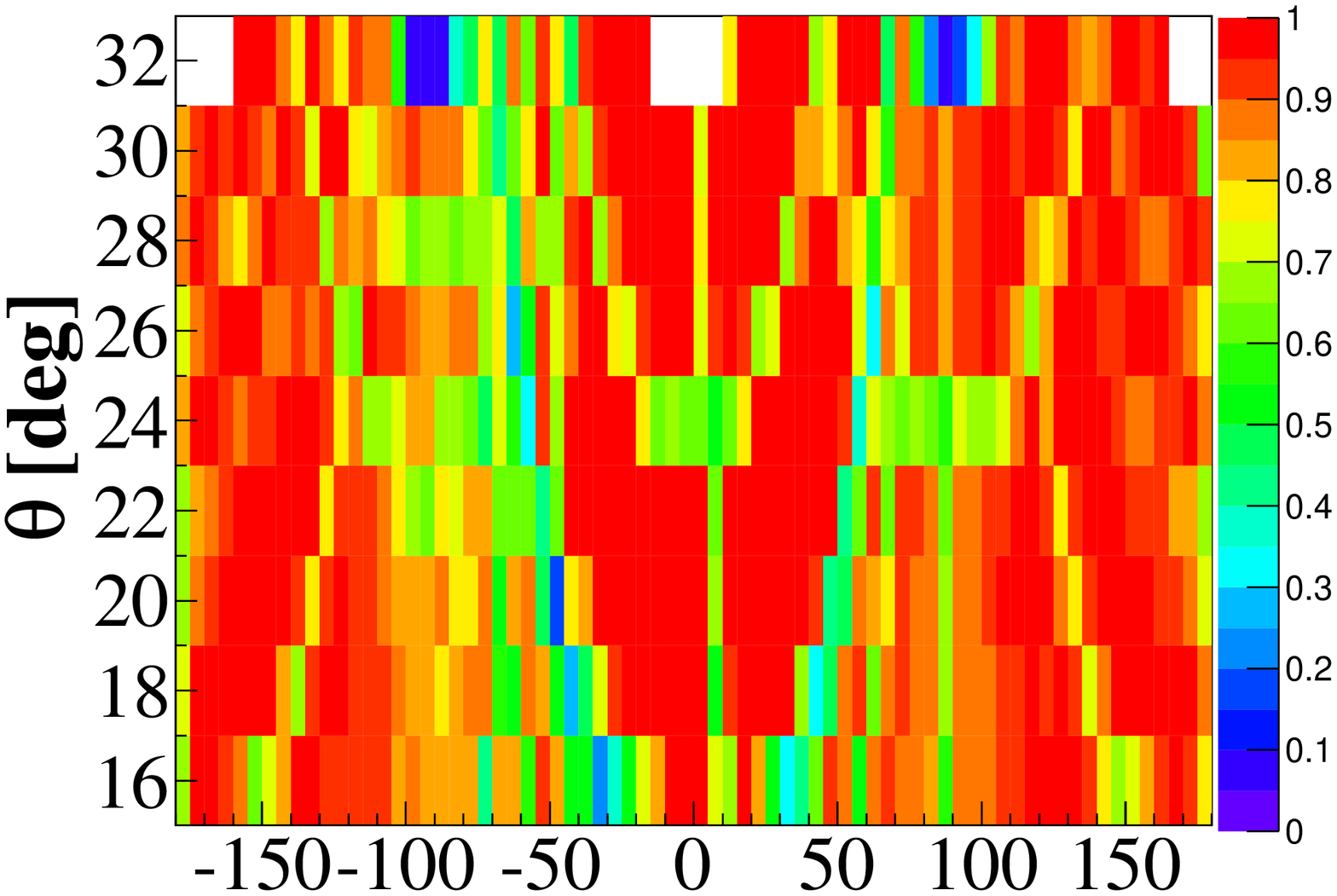}
 \includegraphics[width=0.38\textwidth]{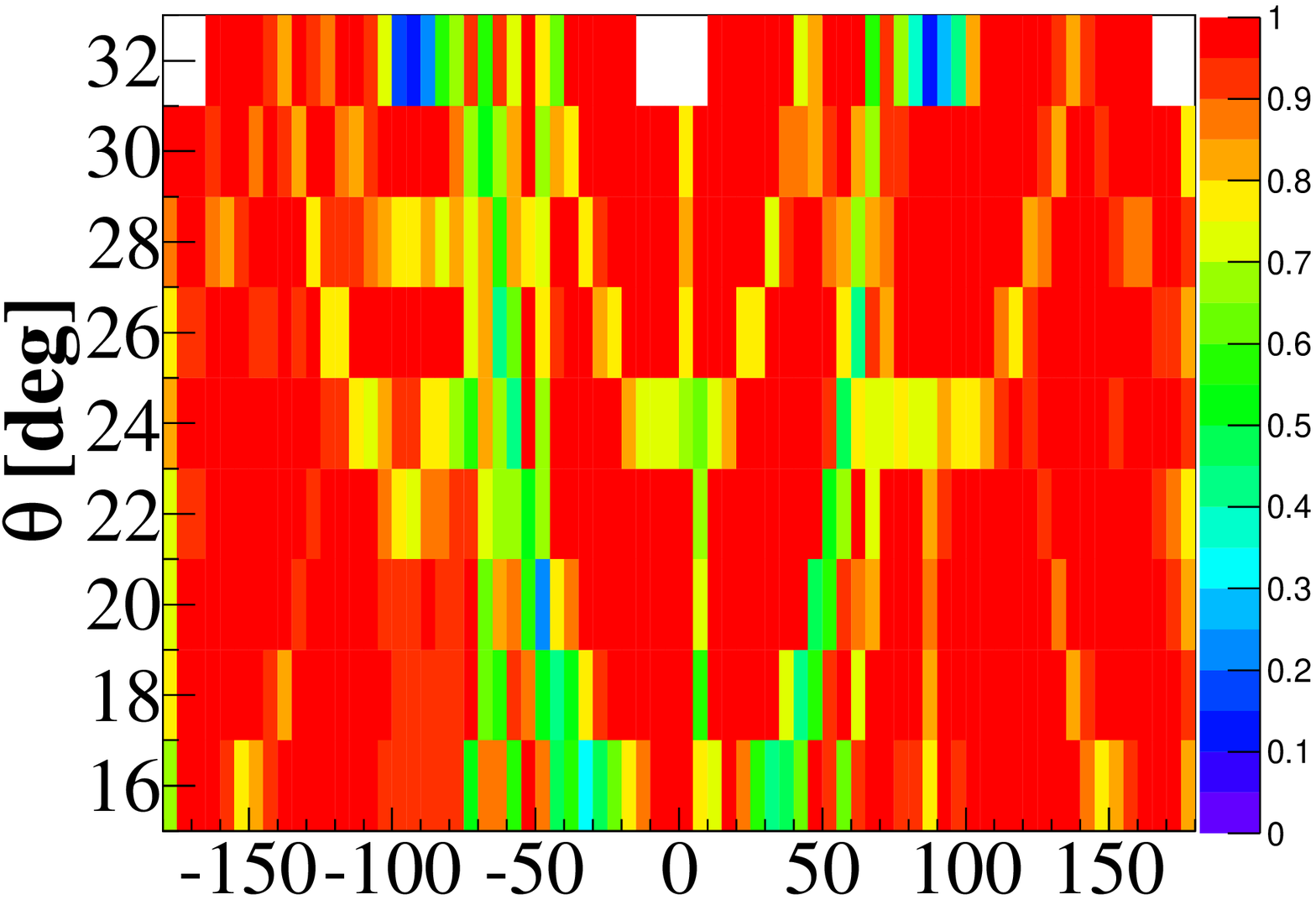}
 \includegraphics[width=0.38\textwidth]{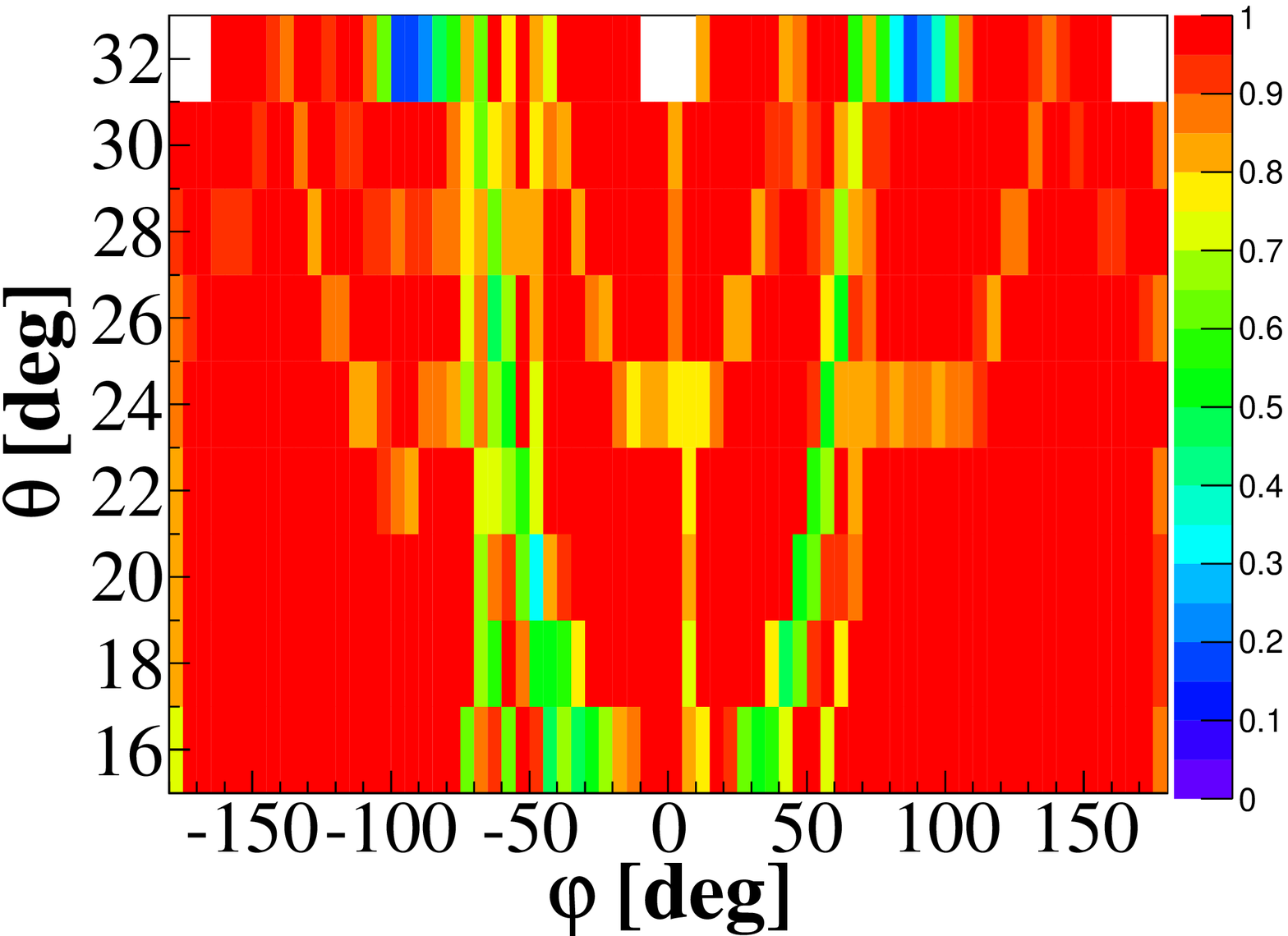}
 \caption{(Color online) The efficiency maps of MWPC for the three $E_{loss}$ ranges. The upper map refers to
 the lowest values of $E_{loss}$ (the fastest particles), whereas the bottom one is 
 constructed for the highest $E_{loss}$ (the slowest particles).
 The elliptic-like structures correspond to ``dead`` or malfunctioning wires.}
 \label{fig3abc}
 \end{figure}
\begin{eqnarray}
\varepsilon_{xyu}^{weak} & = &\varepsilon_{xyu}+\varepsilon_{x}\cdot\varepsilon_{y}\cdot(1-\varepsilon_{u})+\varepsilon_{y}\cdot\varepsilon_{u}
\cdot(1-\varepsilon_{x}) \nonumber\\
&&+\varepsilon_{u}\cdot\varepsilon_{x}\cdot(1-\varepsilon_{y}).
\label{eq2a}
\end{eqnarray}

\begin{figure}[!h]
  \includegraphics[width=0.35\textwidth]{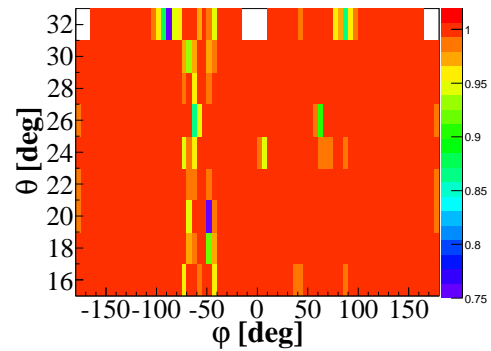}
 \caption{(Color online) The MWPC efficiency map calculated for weak tracks.}
 \label{fig3b1}
 \end{figure} 
 The resulting efficiency map is presented in Fig. \ref{fig3b1}.
 The total MWPC efficiency for weak tracks is about 98\%.\\
\indent  On the analogy to the case of MWPC the efficiency map
 of the $\Delta E$ detector was also constructed. 
 The accumulated inefficiency of this detector was below 5\%.  
 The efficiency of the {\em E} detector has been assumed to be 100\%.  
 \subsubsection{Configurational efficiency}
 \label{effic2}
\begin{figure}[!h]
 \includegraphics[width=0.45\textwidth]{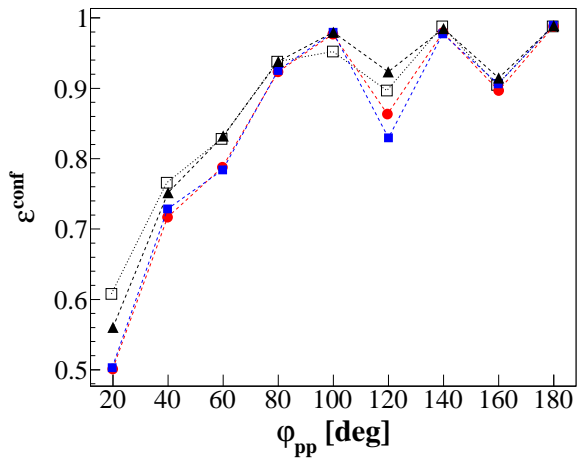}
\caption{(Color online) Configurational efficiency $\varepsilon^{conf}$ for proton-proton coincidences (see Eq. \ref{eq3})
calculated for $\theta_{p1}=19^{\circ}$, $\theta_{p2}=17^{\circ}$ in three cases: $+\varphi_{pp}$ (blue squares),  $-\varphi_{pp}$ (black triangles) 
and $\pm\varphi_{pp}$ (red circles). Empty squares represent results from the simulations. 
Lines connecting points are used to guide the eye.}
 \label{fig3d}
 \end{figure}

\begin{figure}[!h]
 \includegraphics[width=0.45\textwidth]{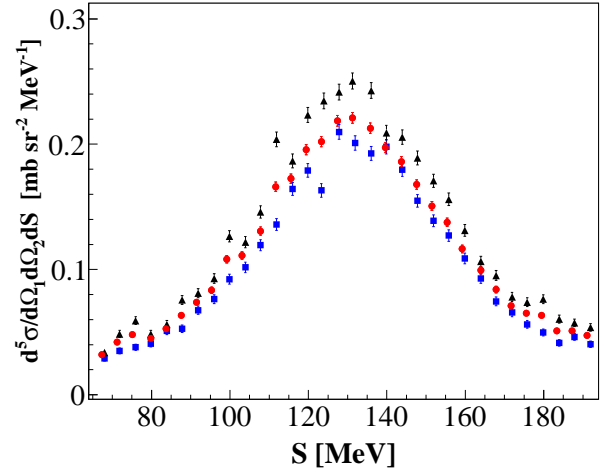}
 \includegraphics[width=0.5\textwidth]{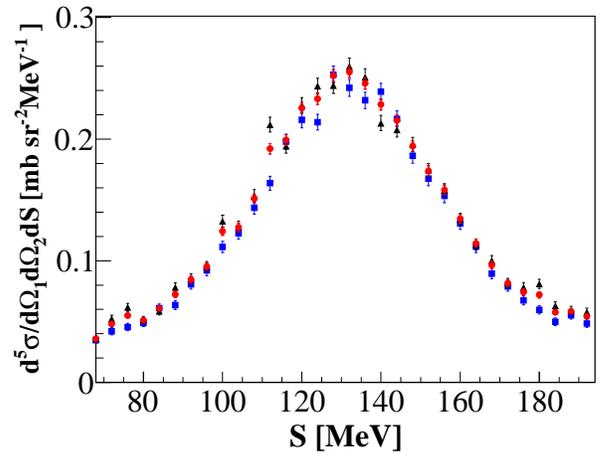}
\caption{(Color online) Cross section distributions obtained for configuration 
$\theta_{p1}=19^{\circ}$, $\theta_{p2}=17^{\circ}$, $\varphi_{pp}=120^{\circ}$
before (upper panel) and after (lower panel) correction for configurational efficiency.
Cross sections are presented for three cases for $+\varphi_{pp}$ (blue squares),  $-\varphi_{pp}$ (black triangles) 
and $\pm\varphi_{pp}$ (red circles).}
 \label{fig3e}
 \end{figure}

\indent Due to the fact that the events of interest are coincidences of the two particles
registered in {\em E} and $\Delta E$ detectors of a finite granulation, additional losses
of acceptance have to be taken into account. The events with both particles registered in the same $\Delta E$
or {\em E} detector have to be rejected due to the lack of particle identification and/or unknown energy, see also Sec.~\ref{data1}.
Since the $\Delta E$ detector has two times higher granulation than {\em E} detector (24 scintillators compared to 10)
this effect is dominated by the {\em E} detector. 
Such inefficiency, in the following referred to as the configurational efficiency, 
is of geometrical origin and relevant only for coincidences. It is expected to be  
pronounced at low relative azimuthal angles of the proton-deuteron pair ($\varphi_{dp}\leq 80^{\circ}$).\\
\indent To establish this efficiency the data collected with 160 MeV deuteron beam impinging on the proton target (\cite{Parol:17}) 
were used. In the {\em dp} scattering only two channels are present: elastic scattering and {\em ppn} deuteron breakup,
in contrary to the {\em dd} scattering \cite{Cie:18}, where also 4-body breakup and transfer channels contribute. 
Therefore, any non-coplanar configuration of charged particles in {\em dp} scattering data 
can be interpreted as the {\em ppn} breakup or, very unlikely, an accidental coincidence.   
The {\em ppn} channel has also advantage of low cross section at the edges of kinematical curves. 
In contrary to the {\em dpn} channel dominated by the quasi-free process in which the events are usually gathered on the 
edges of the kinamatical curves (near detection thresholds) which is important in view of the discussion below. 
The configurational efficiency (see Sec.~\ref{cs} for definition of configuration) was 
calculated with the use of the {\em particular tracks} (defined in Sec.~\ref{data1}) for   
each geometrical configuration ($\theta_{p1}$, $\theta_{p2}$, $\varphi_{pp}$) analyzed in this paper with the same angular 
bins of $\Delta\theta=2^{\circ}$, $\Delta\varphi_{pp}=10^{\circ}$ as applied in the analysis of the {\em dpn} breakup.
The particular tracks defined in Sec.~\ref{data1} include also events with one particle 
stopped in $\Delta E$ or with small energy deposit in {\em E}-detector (below the threshold). 
In order to reject such events or minimize their impact, an upper limit was set on the energy deposited in $\Delta E$.
 Based on the {\em ppn} breakup, angular information from MWPC (no PID available) and energy 
 deposited in the $\Delta E$ detector the efficiency was constructed as follows:
\begin{eqnarray}
&\varepsilon^{conf}(\theta_{p1}, \theta_{p2}, \varphi_{pp})=
\frac{N_{break}(\theta_{p1}, \theta_{p2}, \varphi_{pp})}{N_{ce}(\theta_{p1}, \theta_{p2}, \varphi_{pp}) + N_{break}(\theta_{p1}, \theta_{p2}, \varphi_{pp})},
\label{eq3}
\end{eqnarray}
where $N_{ce}(\theta_{p1}, \theta_{p2}, \varphi_{pp})$ denotes 
the number of the {\em p}-{\em p} coincidences registered as the \textit{particular tracks},
whereas $N_{break}(\theta_{p1}, \theta_{p2}, \varphi_{12})$ denotes the number of the coincidences for the complete tracks. 
$N_{ce}(\theta_{p1}, \theta_{p2}, \varphi_{pp})$ and $N_{break}(\theta_{p1}, \theta_{p2}, \varphi_{pp})$
represent experimental values obtained by integrating the events over the arclength {\em S}.
 In order to check our method, the data were analyzed for 
 the kinematical configurations with $+\varphi_{pp}$ (so-called normal configurations) 
 and $-\varphi_{pp}$ (so-called mirror configurations) separately, and for $\pm\varphi_{pp}$ (normal and mirror configurations treated together).
 As an example the configurational efficiency for $\theta_{p}=19^{\circ}$, $\theta_{p}=17^{\circ}$ is presented, see Fig.~\ref{fig3d}. 
 In the case of $\varphi_{pp}=120^{\circ}$, values of $\varepsilon^{conf}$ differ significantly between normal and mirror configurations. 
Therefore, the corrections should lead to the same results for the corresponding cross sections
for the normal and mirror configurations, as follows from the parity conservation.
In Fig.~\ref{fig3e} the cross sections before (upper panel) and after (lower panel) the efficiency corrections are presented. 
The corrected cross sections are consistent within statistical errors.  
The corrections based on experimental data were confronted with the GEANT4 simulations.
In the simulations, the uniform three-body breakup phase-space distribution 
has been used, which is well-justified in the case of narrow angular ranges applied in defining the
configuration. The number of the breakup coincidences were counted and, simultaneously, 
fraction of the breakup coincidences with two protons registered in the same {\em E} or $\Delta E$ detector, was obtained. 
In Fig.~\ref{fig3d} samples of $\varepsilon^{conf}$ are presented for 
the data and the simulations.
The configurational efficiency calculated with the use of the simulations can be considered
as purely geometrical factors
of probability for double hits in single-{\em E}-bar or single-$\Delta E$-strip events. 
The data and the simulations agree qualitatively and usually the differences vary between 2-4\% for the range 
of $\varphi_{dp}$ above 120$^{\circ}$, discussed in this paper. The statistical uncertainties are within 0.1. 
These differences are due to the fact that the simulations are simplified.
They do not include data digitization and also the virtual detector geometry 
is not ideally modeled. Therefore, the further analysis relays on the corrections obtained directly from the experimental data.\\
\indent In the range of $\varphi_{dp}$ between $140^{\circ}$ and $180^{\circ}$, which is being examined, 
practically only the efficiency losses due to double hits in the {\em E}-detector matter.
For the $\Delta E$ and MWPC detectors the configurational inefficiencies are negligible (below 1\%).
\subsection{Hadronic interactions}
\label{hi}
\begin{figure}[!h]
\centering
\includegraphics[width=0.5\textwidth]{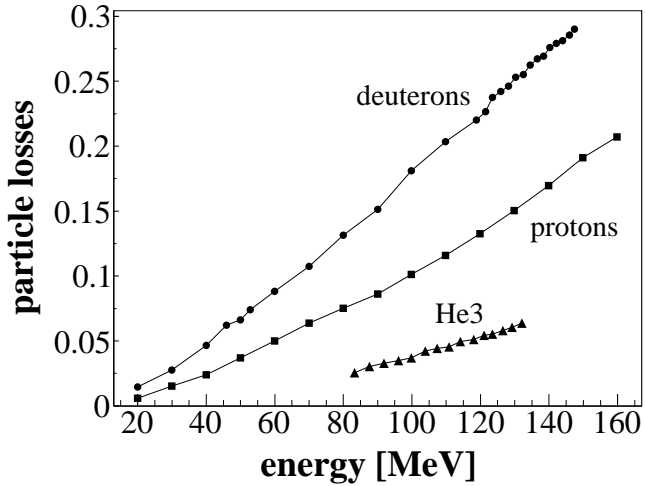}
\caption{Results of simulation of relative loss of events due to hadronic 
interaction of particles stopped in the plastic scintillator. The losses for 
protons, deuterons and $^{3}$He-ions are presented as a function of initial energy. }
\label{fig4a}
\end{figure}

Additional losses of events took place due to hadronic interactions inside the scintillator material.
They were treated as a background (see Sec.~\ref{cs}) together with events induced by charged particles on passive material of the detector setup. 
The losses were calculated for the protons and deuterons in the energy range of interest with the use of the GEANT4 framework \cite{Cie:18}.
The results of the calculations are presented in Fig.~\ref{fig4a}. 

\subsection{Quasi-free scattering}
\label{qfs}
\begin{figure}[!h]
\centering
\includegraphics[width=0.5\textwidth]{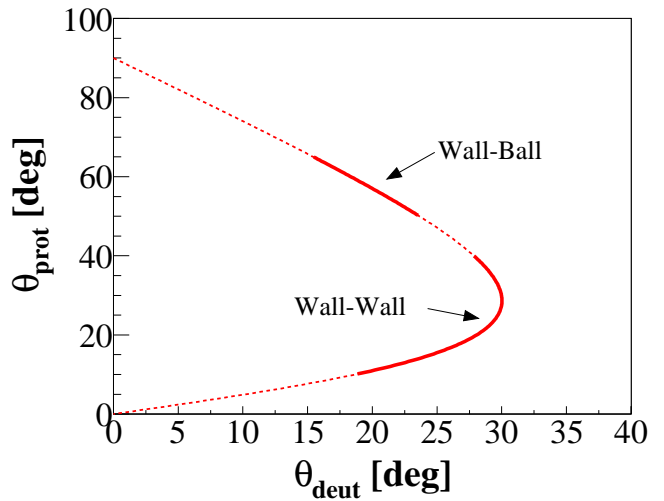}
\caption{Kinematic relation of the {\em dp} elastic scattering at 157.7 MeV.}
\label{fig4b}
\end{figure}

The quasi-free {\em dd} scattering (QFS) occurs when one of the colliding deuterons 
knocks-out the proton of another deuteron 
in kinematic conditions close to the free scattering. 
Momentum of the accompanying neutron is not changed so, in a simplified approach, 
the neutron can be treated as a spectator of the reaction. 
A deuteron binding energy of 2.224 MeV is negligible
in comparison to the beam energy of 160 MeV and  
in this case the process is dominated by the interaction with a single nucleon.
The QFS kinematics can be realized in two ways; \textit{(i)} the beam deuteron is scattered on proton of the deuteron target ({\em dp}-QFS),
and \textit{(ii)} the proton of the beam deuteron is scattered on deuteron
target ({\em pd}-QFS). For the {\em dp}-QFS, the reaction energy is about 157.7~MeV, 
while in the latter case about 77.7~MeV (about half the beam energy). 
In this paper only the {\em dp}-QFS is considered 
since in this case both outgoing charged particles can be registered within 
the forward Wall acceptance. Fig. \ref{fig4b} shows
the kinematical relations for the {\em dp}-QFS scattering in a situation when the neutron
spectator is at rest in the laboratory frame.

\subsection{Breakup cross section}
\label{cs}
\begin{figure}[!h]
\centering
\includegraphics[width=0.45\textwidth]{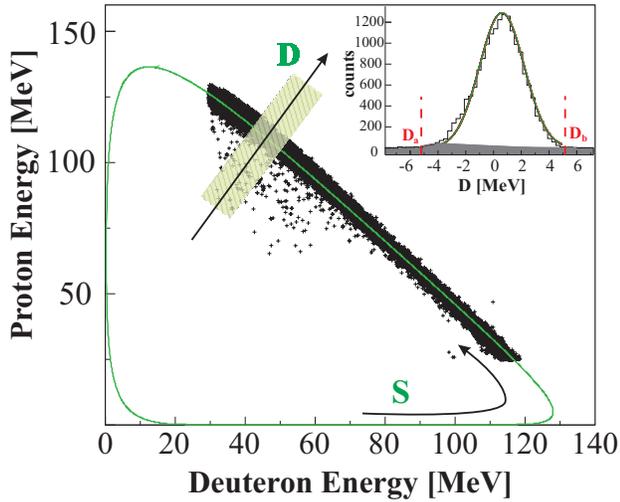}
\caption{(Color online) $E_{p}$ $vs.$ $E_{d}$ coincidence spectrum of the proton-deuteron pairs registered
at $\theta_{d}=20^{\circ}\pm1^{\circ}$, $\theta_{p}=18^{\circ}\pm1^{\circ}$, $\varphi_{dp}=160^{\circ}\pm5^{\circ}$. 
The solid line shows a 3-body kinematical curve calculated for the central values of the angular ranges. Variables 
arc-length {\em S} and distance from kinematics {\em D} are presented in a schematic way.
The inset presents the {\em D} distribution of events belonging to one $\Delta S$ bin. 
The Gaussian distribution was fitted in the range of {\em D} corresponding
to distances of $D_{a}$=-3$\sigma$ and $D_{b}$=+3$\sigma$ from the fitted peak position.}
\label{fig5}
\end{figure}

The geometry of a coincident proton-deuteron ({\em p}-{\em d}) pair is characterized by their polar
angles $\theta_{d}$ and $\theta_{p}$ and relative azimuthal angle $\varphi_{dp}$. 
When two particles are measured in coincidence at a
particular pair of angles the particle energies are fully
determined. Momentum and energy conservation and the relation $\varphi_{dp}=\mid\varphi_{d}-\varphi_{p}\mid$
unambiguously define the kinematics of the three-body breakup which is described with 
five independent variables $E_{d}$, $E_{p}$, $\theta_{d}$, $\theta_{p}$ and $\varphi_{dp}$.
The relation between energies $E_{p}~ vs. ~E_{d}$ is represented with the kinematical curve, see Fig.~\ref{fig5}.
The energies $E_{p}$ and $E_{d}$ were transformed into two new variables  
in the $E_{p}~ vs. ~E_{d}$ plane (see Fig.~\ref{fig5}): {\em D} denoting 
the distance of the $(E_{d}, E_{p})$ point from the kinematical curve for the point-like geometry, 
and {\em S}, which defines the arc-length along the kinematics with the starting point at the minimal $E_{p}$.
In the analysis, the angular ranges for kinematic spectra were chosen as 
follows: $\Delta\theta_{d}=\Delta\theta_{p}=2^{\circ}$, $\Delta\varphi_{dp}=10^{\circ}$ and 
are wide enough to reach a good statistical accuracy. 
The events in each bin of $\Delta S$~=~4~MeV (see Fig.~\ref{fig5}), corrected previously for efficiencies, 
were projected onto the {\em D}-axis. The sample distribution in a function of the variable {\em D} 
is presented in the inset of Fig.~\ref{fig5}.
The breakup events are grouped in a prominent peak with only a very low background.
Since the exact shape of the background is not known, as the
first approximation, linear behavior was assumed. 
To calculate the cross section in a function of {\em S}, 
the Gauss function was fitted to the {\em D}-distributions. To treat all configurations
in a consistent way, the integration limits in {\em D} variable were chosen 
at the values of $D_{a}$ and $D_{b}$, see the inset of Fig.~\ref{fig5}, 
corresponding to distances of -3$\sigma$ and +3$\sigma$ from the maximum of
the fitted peak.\\
\indent Measurements with an unpolarized beam and a detector with axial 
symmetry allows for integration of events over polar angles. 
Number of the proton-deuteron breakup coincidences $N_{br}(S, \Omega_{d}, \Omega_{p})$ registered 
at given angles $\Omega_{d}\equiv~(\theta_{d},\varphi_{d})$ and $\Omega_{p}\equiv(\theta_{p},\varphi_{p}=\varphi_{d} + \varphi_{dp})$ 
and in a {\em S} arc-length bin, is given as follows: 
\begin{eqnarray}
N_{br}(S,\Omega_{d},\Omega_{p})&=&\frac{d^{5}\sigma}{d\Omega_{d}d\Omega_{p}dS}(S,\theta_{d},\theta_{p},\varphi_{dp}) \nonumber\\
& &\times \kappa \cdot \Delta \Omega_{d} \Delta\Omega_{p} \Delta S \nonumber\\
& &\times \varepsilon^{d}_{xyu}(\theta_{d},\varphi_{d},E_{loss}) \cdot \varepsilon^{p}_{xyu}(\theta_{p},\varphi_{p},E_{loss}) \nonumber\\
& &\times \varepsilon^{d}_{\Delta E}(\theta_{d},\varphi_{d}) \cdot \varepsilon^{p}_{\Delta E}(\theta_{p},\varphi_{p}) \nonumber\\
& &\times \varepsilon^{conf}(\theta_{d},\theta_{p},\varphi_{dp}),
\label{eq1}
\end{eqnarray}
where $\frac{d^{5}\sigma}{d\Omega_{d}d\Omega_{p}dS}$ denotes differential cross section for the breakup reaction 
for a chosen angular configuration; solid angles are calculated as $\Delta \Omega_{j}=\Delta \theta_{j}\Delta \varphi_{j}$sin$\theta_{j}$,
{\em j=~d,~p}. $\varepsilon^{d}_{xyu}(\theta_{d},\varphi_{d},E_{loss})$ and $\varepsilon^{p}_{xyu}(\theta_{p},\varphi_{p},E_{loss})$
are the MWPC efficiencies, whereas 
$\varepsilon^{d}_{\Delta E}(\theta_{d},\varphi_{d})$ and $\varepsilon^{p}_{\Delta E}(\theta_{p},\varphi_{p})$
are {\em dE} efficiencies for deuteron and proton, respectively. 
$\varepsilon^{conf}(\theta_{d},\theta_{p},\varphi_{dp})$ is the configurational efficiency (see Sec. \ref{effic2}). 
$\kappa$ is the normalization factor defined in the next Section ~\ref{norm}.

\subsection{Cross section normalization}
\label{norm}
The differential cross sections for the $^{2}$H($d$,$dp$)$n$ breakup reaction were normalized to the 
known {\em d}-{\em d} elastic scattering cross-section data. For that purpose the so-called scaling 
region characterized with very weak energy dependence of the cross section was used \cite{Cie:18}, 
which enables scaling of the measured elastic scattering rate to the data 
at two closest energies (130 and 180 MeV) \cite{Bail:09}.
Such a relative normalization method ensures cancellation of factors related to the luminosity (i.e., the beam
current, the density or the thickness of the target) or to the electronic and readout dead-times.
In this way, we profit from cancellation of many factors which are hard to determine and can be a source
of systematic uncertainties.
The procedure of extracting the normalization factor $\kappa$,
which corresponds to the luminosity integrated over the time of the data collection, is described in details in \cite{Cie:18}.
For the breakup cross section $\kappa$ was established to be $48.4\pm3.9$ (syst.)*10$^{6}$~mb$^{-1}$.
\subsection{Experimental uncertainties}
\label{expun}
\begin{figure}[!th]
 \includegraphics[width=0.4\textwidth]{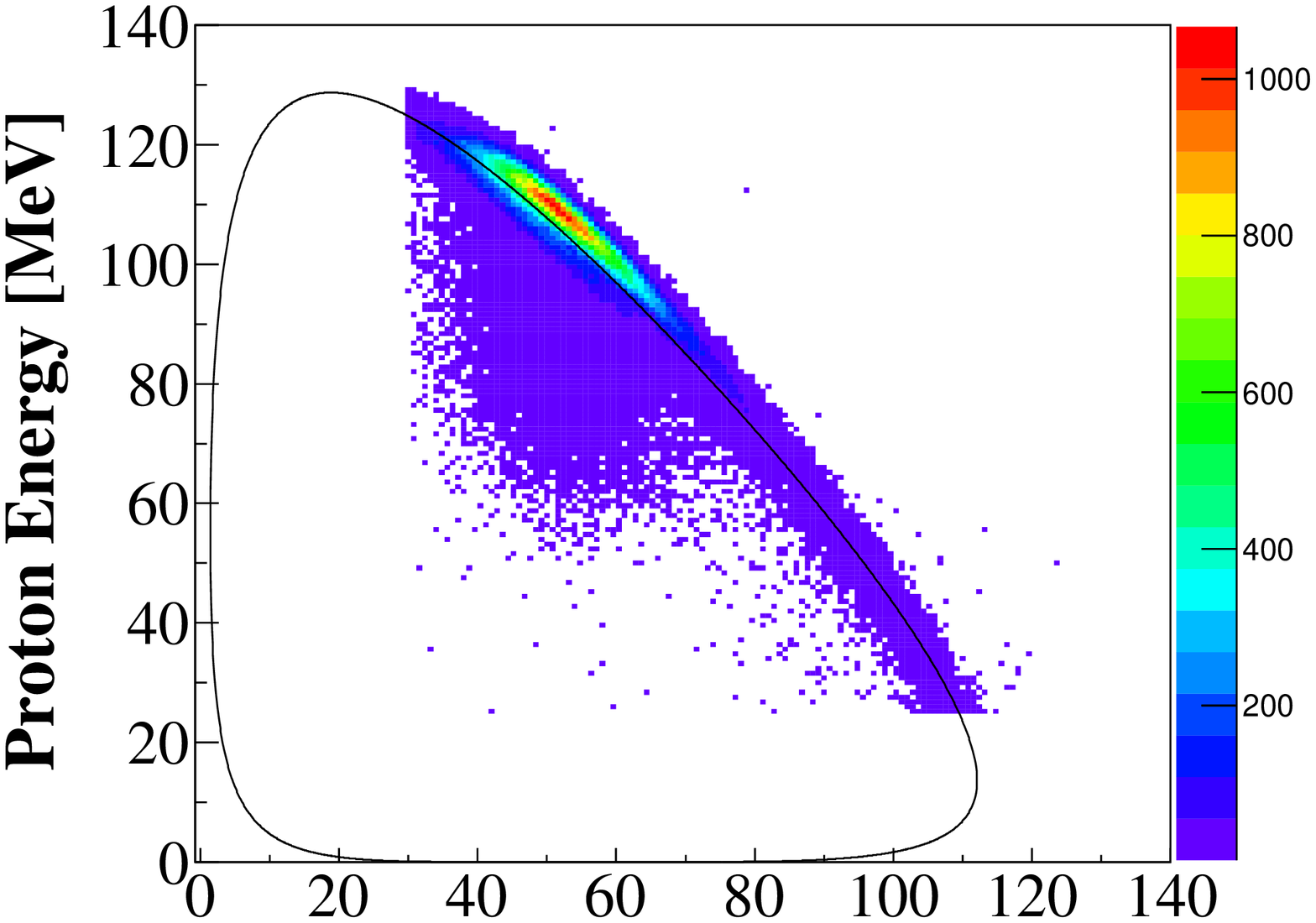}
 \includegraphics[width=0.4\textwidth]{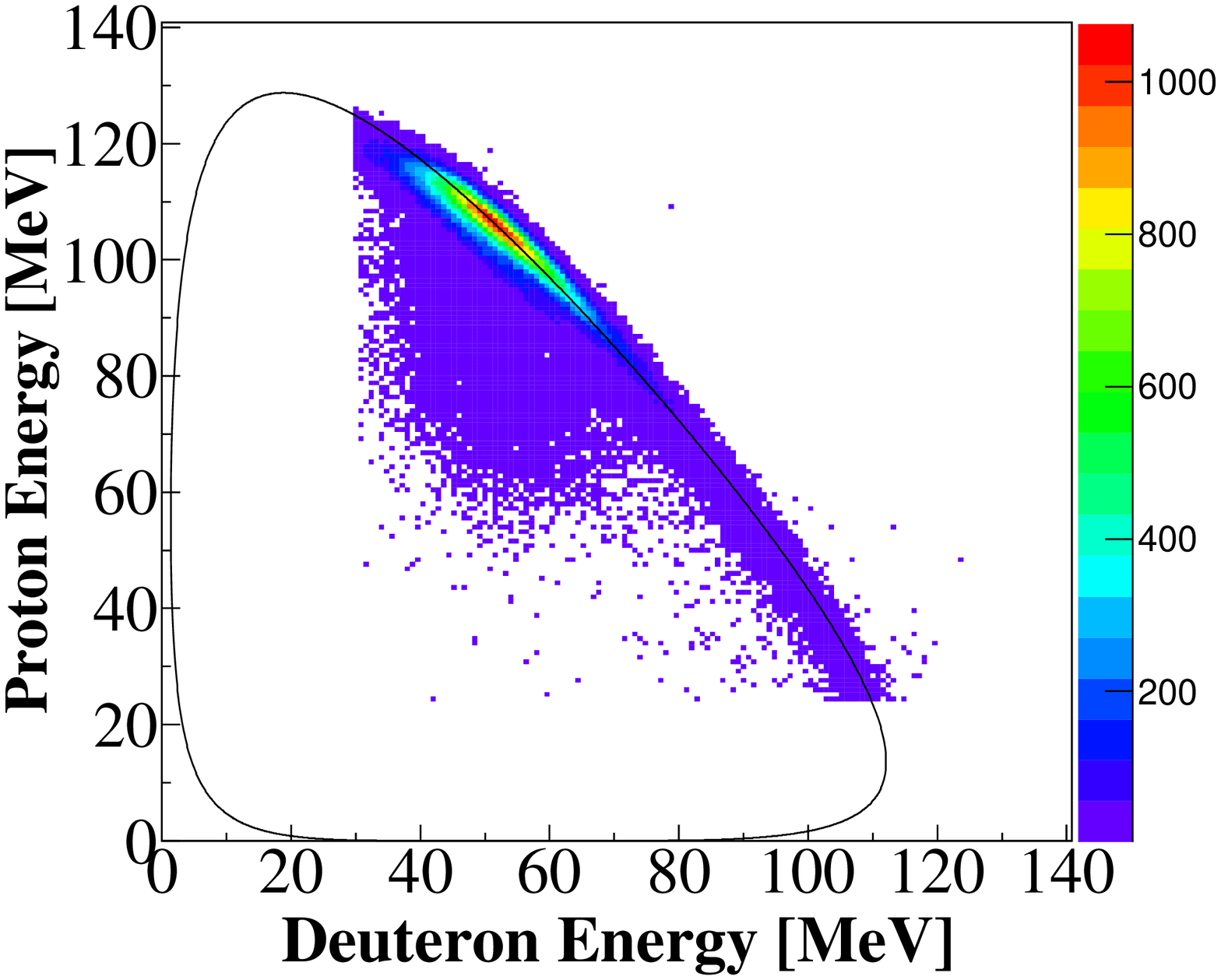}
 \includegraphics[width=0.4\textwidth]{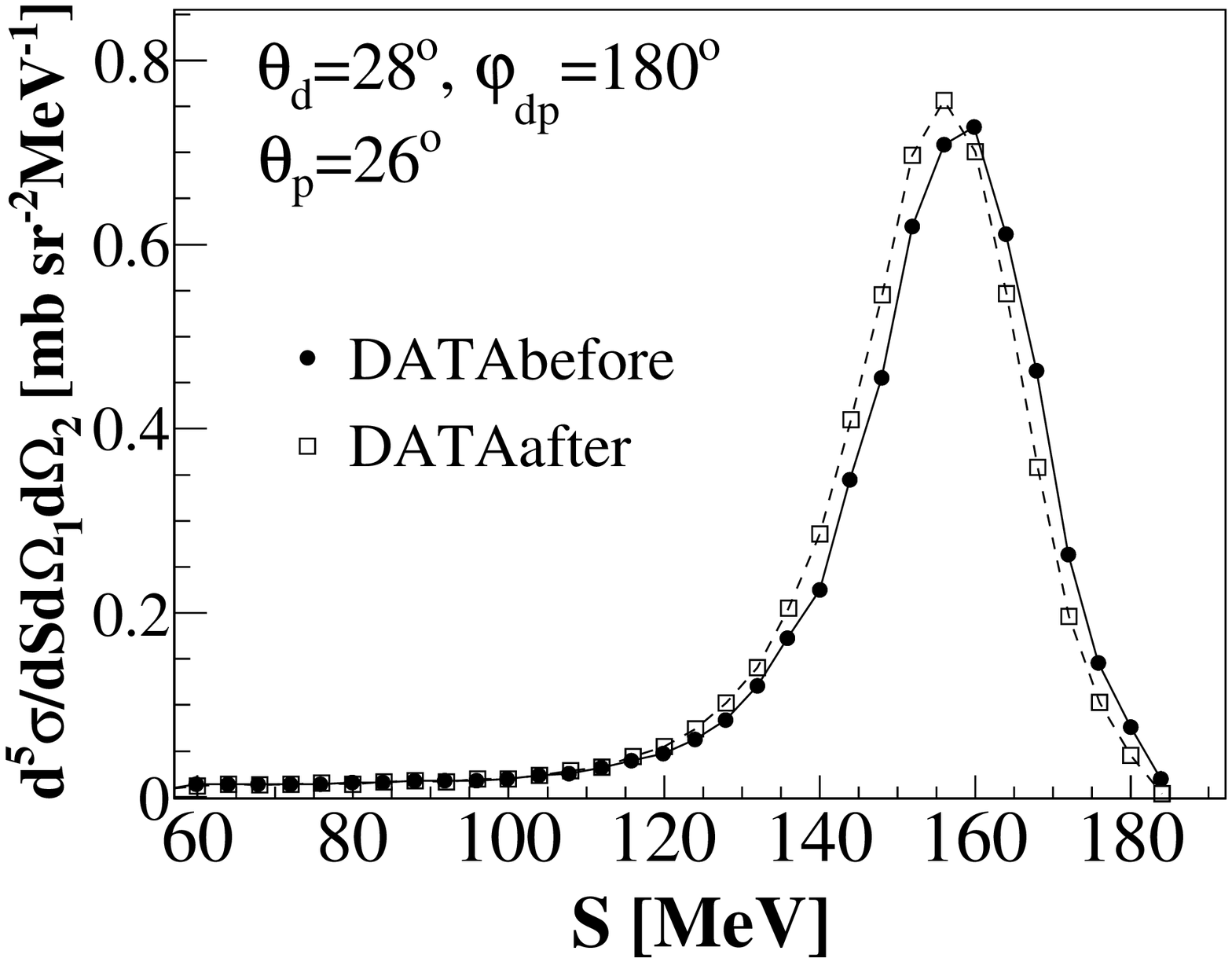}
 \caption{(Color online) $E_{p}$ vs. $E_{d}$ kinematical relation for 
 the $\theta_{d}=28^{\circ},~\theta_{d}=26^{\circ},~\varphi_{dp}=180^{\circ}$ configuration drawn together with the theoretical
 kinematical curve (\textit{upper panel}). \textit{Middle panel:} the experimental kinematics scaled to the theoretical relation.
 \textit{Bottom panel:} The comparison of the cross section distributions obtained for the above two cases demonstrating the size
 of the systematic effect accounted for the calibration procedure. Lines connecting points are used to guide the eye.}
 \label{fig30b}
 \end{figure}

The most serious sources of the systematic uncertainties which affect the breakup cross section 
are related to the PID method, normalization procedure, data averaging effect and the track reconstruction procedure.
The systematic effects related to the normalization procedure were already described in \cite{Cie:18}, here the
systematic effects are discussed in the context of the breakup cross section. \\
\indent Protons and deuterons were identified via defining graphical cuts 
enclosing the branches on the $\Delta E$-{\em E} spectra. 
A finite precision in defining such cuts may lead to mixing of particle
types, or cutting out a part of useful events. The systematic uncertainty 
associated with this process was estimated by repeating the analysis based on broaden
and narrowed cuts of around 0.5$\sigma$ of the original cut.
Based on this the relative difference of the resulting
cross sections were calculated.
The typical uncertainty of the final
breakup cross section related to this effect do not exceed 5\%. \\
\indent The amount of background, which is caused by the hadronic interactions (see Sec.~\ref{hi})
and to a lesser extent by the accidental coincidences 
was found low, so the uncertainty related to background subtraction can be neglected (see Fig.~\ref{fig5}, inset). 
However, for a few configurations and certain {\em S}-bins the background contribution was significant and 
systematic effects connected to the background subtraction was found to be 2\%-5\%.\\
\indent For some sets of configurations significant systematic effects related to the calibration
were observed. For these geometries shape of the experimental kinematical curves differs from the
theoretical one which influences the distributions of the cross section. 
In Fig.~\ref{fig30b} example for the $\theta_{d}=28^{\circ},~\theta_{d}=26^{\circ},~\varphi_{dp}=180^{\circ}$ configuration 
is presented: upper and middle panels show the original and scaled (so that to fit to the theoretical curve) 
distributions, respectively. The bottom figure presents the resulting cross-section distributions. 
The size of the effect
related to the events migration between different {\em S}-bins was estimated to be 3\%-10\%.\\
\indent The experimental cross section for a given angular configuration ($\theta_{d}$, $\theta_{p}$, $\varphi_{dp}$)
is evaluated by taking a finite bin width around these angles, i.e. $\theta_{d}\pm\frac{1}{2}\Delta\theta_{d}$,
$\theta_{p}\pm\frac{1}{2}\Delta\theta_{p}$ and $\varphi_{dp}\pm\frac{1}{2}\Delta\varphi_{dp}$.
\begin{figure}[!h]
\centering
\includegraphics[width=0.4\textwidth]{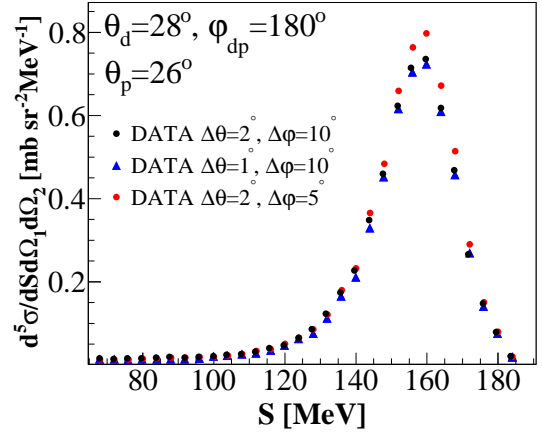}
\caption{(Color online) Comparison of sample cross sections obtained with a regular bin size of
$\Delta\varphi_{dp}$=10$^{\circ}$, smaller bin size of $\Delta\varphi_{dp}$=5$^{\circ}$
and $\Delta\theta_{p}$=1$^{\circ}$.}
\label{fig30}
\end{figure}
The bin width is taken wide enough (here $\Delta\theta_{d}$=$\Delta\theta_{p}$=2$^{\circ}$ 
and $\Delta\varphi_{dp}$=10$^{\circ}$) to assure good statistical accuracy. On the
other hand, the theoretical predictions, used for the comparison, were calculated
at the central values of these angular bins. As explained in Ref.~\cite{Cie:15}, 
the averaging of the calculations over the experimental bin width is quite crucial 
to validate theories in an reliable way.
In order to estimate the associated systematic error, related to the size of the angular
bin width the analysis was performed with smaller $\Delta\varphi_{dp}$ bin size of 5$^{\circ}$
and $\Delta\theta_{p}$=1$^{\circ}$, see Fig.~\ref{fig30}. 
The systematic error 
associated with this effect, in most cases, was found to be up to~5\%.\\
\indent Additional loss of events is related to the so-called crossover events \cite{Kis:05}.
Such events occur when particles penetrate from one stopping detector to the adjacent one and in this case events are lost due to 
distorted energy information. In this experiment due to improper light tightness between
{\em E}-slabs uncontrolled light leakage increased the crossovers. 
The systematic effects originating from the track reconstruction procedure and crossover losses were 
estimated based on the three datasets obtained separately for complete and weak tracks and with taking into account crossovers events in
the complete tracks reconstruction, see Sec. \ref{data1}, \ref{expun}.
Such errors were calculated for each individual configuration.
\begin{table}[H]
 \centering
\caption{Sources of systematic effects and their influence (in \%) on the 
  breakup cross section.}
 \label{tab1}
 \begin{tabular}{|c|c|}
 \hline
 \textbf{Source of uncertainty} & \textbf{Size of the effect} \\\hline
 PID & 5\% \\\hline
 Normalization & 8\% (\cite{Cie:18}) \\\hline
 Reconstruction of angles & 1\% \\\hline
 Energy calibration & 1\% (maximum 10\%) \\\hline
 Background subtraction & 1\% (maximum 5\%) \\\hline
 Averaging effect & 5\% \\\hline
 Configurational efficiency & maximum 5\%\\
 &(for selected configurations) \\\hline
 Cross-section spread & 7\% (maximum 12\%) \\\hline
 \textbf{TOTAL} & \textbf{13\%}(20\%) \\\hline
   \end{tabular}
 \end{table}
The deviations usually reach up to 7\% and the maximum one is around 12\%.\\ 
The systematic errors are depicted as a red band in Figs.~\ref{fig6}-\ref{fig24} presented in Appendix
and are summarized in Table~\ref{tab1}. The total systematic uncertainty composed of systematic
errors added in quadrature varies between 13\% and 20\%.


\section{Results and comparison to the SSA calculations}
\label{results}
The experimental cross-section data were obtained for 147 geometries of the proton-deuteron pairs 
from the breakup reaction near the QFS region, what corresponds to 
relative azimuthal angles $\varphi_{dp}$: $140^{\circ}$, $160^{\circ}$ and $180^{\circ}$. 
Polar angles $\theta_{d}$ and $\theta_{p}$ were varied between $16^{\circ}$ and $28^{\circ}$ with the step of $2^{\circ}$. 
The theoretical predictions, used for the comparison with the experimental cross sections were calculated
at the central values of the chosen angular bins.
\begin{figure}[!h]
\centering
\includegraphics[width=0.45\textwidth]{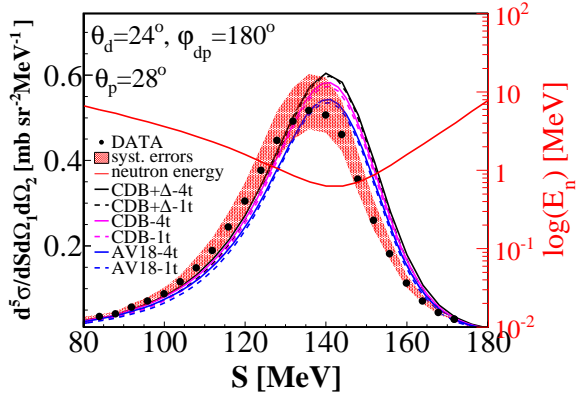}
\caption{(Color online) Example of the differential breakup cross section for the angular configurations
specified in the picture. The experimental points are marked with black dots. The systematic effects 
are depicted as a red band. Various lines represent the theoretical predictions calculated at the central values of 
the defined angular bins.
The black lines refere to the SSA calculations based on the CD Bonn + $\Delta$ (CDB + $\Delta$) potential: solid with {\em 4-term} (4t)
and dashed with {\em 1-term} (1t). Solid and dashed magenta and blue lines represent the similar set of the calculations 
but for the CD Bonn and AV18 potentials, respectively.  
The solid red line present the dependence of the spectator neutron energy (E$_{n}$) along {\em S}-axis.}
\label{fig6}
\end{figure}
Sample cross-section distribution is presented in Fig.~\ref{fig6} 
together with the available SSA calculations, whereas
the results for all individual configurations are collected in Figs.~\ref{fig7}-\ref{fig24} and are presented in Appendix.\\
\indent The SSA calculations are expected to properly estimate the experimental data 
near QFS kinematics (with neutron energy $E_{n}\sim0$) in the center-of-mass (CM) system above 100 MeV. 
High enough beam energy is necessary condition to ensure high 
relative energies for all pairs which is important due to neglected final-state interactions.
With fixed beam energy these conditions correspond to relatively large scattering angles $\theta_{d}$ and $\theta_{p}$.
As it was stated in \cite{Del:16} SSA provides too large cross sections and the discrepancy is decreasing with increasing beam energy. 
The total {\em p}+{\em d} breakup cross section calculated in an exact way is lower than the one obtained in 
SSA by 30\% at 95 MeV and by 20\% at 200 MeV \cite{Del:16}. In {\em Nd} systems, even with $E_{n}\sim0$, 
SSA gives always higher cross sections than the exact calculations.\\
\indent 
The data were sorted according to the relative energy $E_{d-p}$ and the neutron energy $E_{n}$, which in view 
of the above discussion are adequate for validating the SSA results.
The quality of the agreement between the calculations and the experimental cross sections
was studied with the so-called A-deviation factor introduced in \cite{Sha:17} and defined
as follows:\\
\begin{eqnarray}
A\equiv \frac{1}{N}\sum_{i=1}^{N}\frac{\mid\sigma_{i}^{exp}-\sigma_{i}^{th}\mid}{\sigma_{i}^{exp}+\sigma_{i}^{th}},
\label{eq3}
\end{eqnarray}
where the sum runs over number of data points in a given bin of $E_{n}$. 
In the previous analyses of the $ppn$ data \cite{Kis:03,Kis:05,Cie:15,Stephan:10} the standardized $\chi^{2}$ was used 
for the various exact calculations validation, however it is not the case for the $dpn$ breakup.
The current predictions are not yet at the stage to use the $\chi^{2}$ test.
This is due to quite large discrepancies between the data and theories owing to the approximate character of the calculations.
The obvious advantage of the A-factor is its quite simple interpretation \cite{Sha:17}. 
Values of the A-factor belong to the interval [0, 1], where zero means the perfect agreement 
between the data and calculations and with the deterioration of the agreement the A-factor is approaching to one.
Very small values of the A-factor correspond to $\sigma^{th}\approx\sigma^{exp}$
and therefore the A-factor value may be interpreted as a half of the average relative
distance between the experimental and theoretical cross 
\begin{figure}[!bh]
\centering
\includegraphics[width=0.45\textwidth]{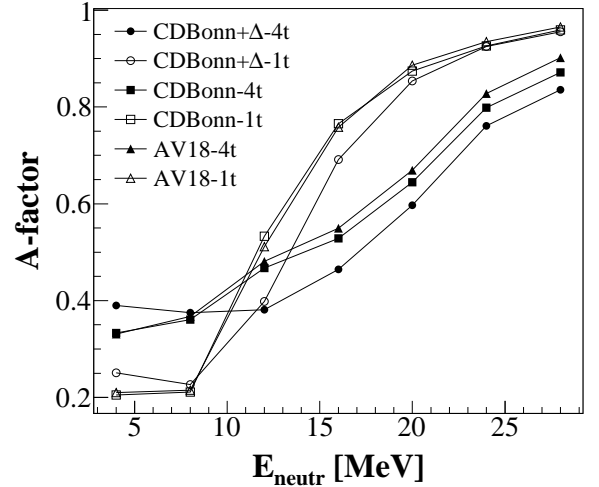}
\caption{Quality of description of the cross-section data with various theoretical predictions (defined in the legend),
expressed as dependence of the A-factor on the neutron energy $E_{n}$. Lines connecting points are used to guide the eye.}
\label{fig29}
\end{figure}
sections.\\
\indent In general one can conclude that at the lowest  $E_{n}$ the {\em 1-term} 
calculations perform better than those with {\em 4-term}, see Fig. \ref{fig29}. 
For low $E_{n}<10$ MeV the {\em 1-term} calculations describe the data well. For higher $E_{n}$
the agreement between the experimental and calculated cross sections deteriorates, but
the {\em 4-term} calculations stay closer to the data.
At highest available $E_{n}$ the A-factor evaluated for all calculations has values close or equal to one 
which means failure of the theoretical description, as expected from the model assumptions \cite{Del:16}.\\
\indent In view of the above discussion on validity of SSA two-dimentional relations $E_{d-p}$ vs. $E_{n}$ were 
also investigated and are presented in Fig. \ref{fig28}, 
with {\em z}-axis representing 
the A-factor for three sets of the calculations: CDB, CDB+$\Delta$ and  AV18 for the {\em 1-} and {\em 4-term} versions. 
Since the A-factor do not account for statistical errors the bins representing poor statistical accuracy are not shown.
In the case of CDB+$\Delta$ the calculations were performed for three relative angles ($140^{\circ}$, $160^{\circ}$ 
and $180^{\circ}$), therefore much more data points contribute, as seen in Fig. \ref{fig28}, middle row. 
For CDB and AV18 the theoretical calculations are available only at $180^{\circ}$.\\
\indent As one can notice the {\em 1-term} predictions better describe the data for $E_{n}<10$,
\begin{figure}[!bh]
\centering
\includegraphics[width=0.5\textwidth]{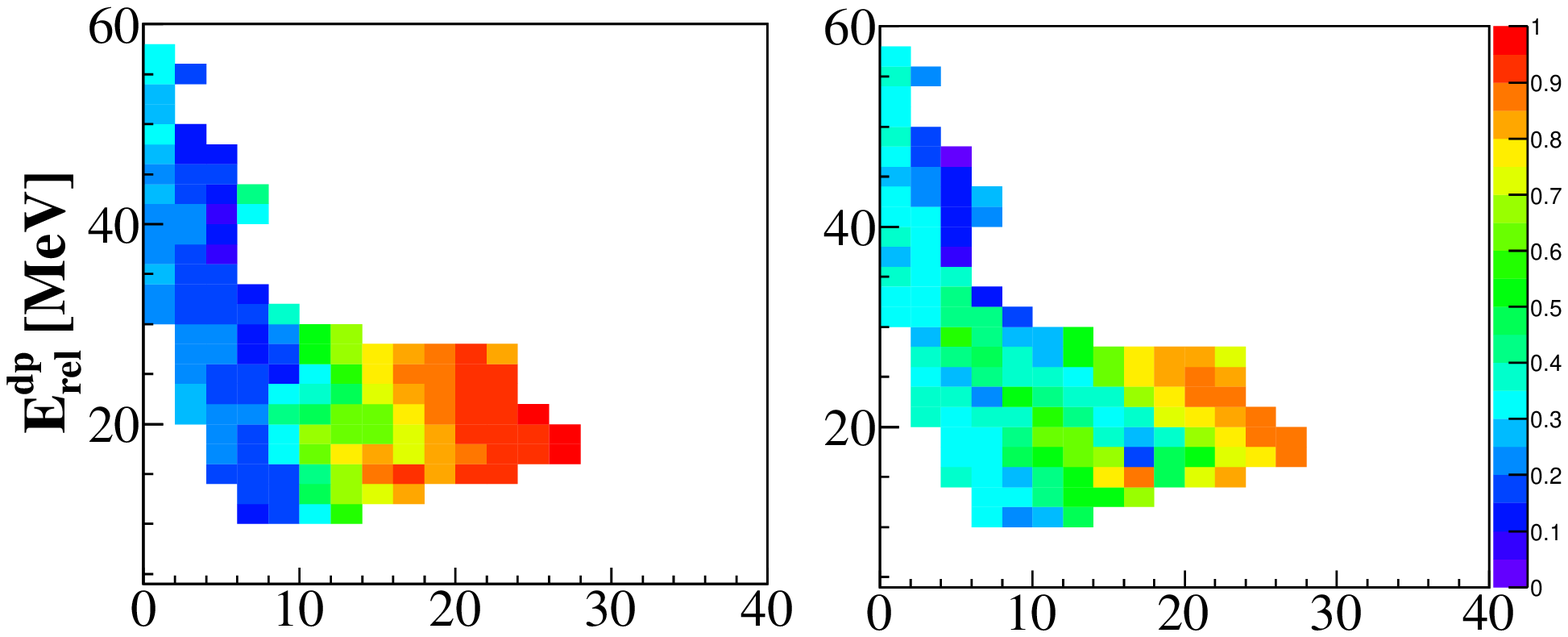}
\includegraphics[width=0.5\textwidth]{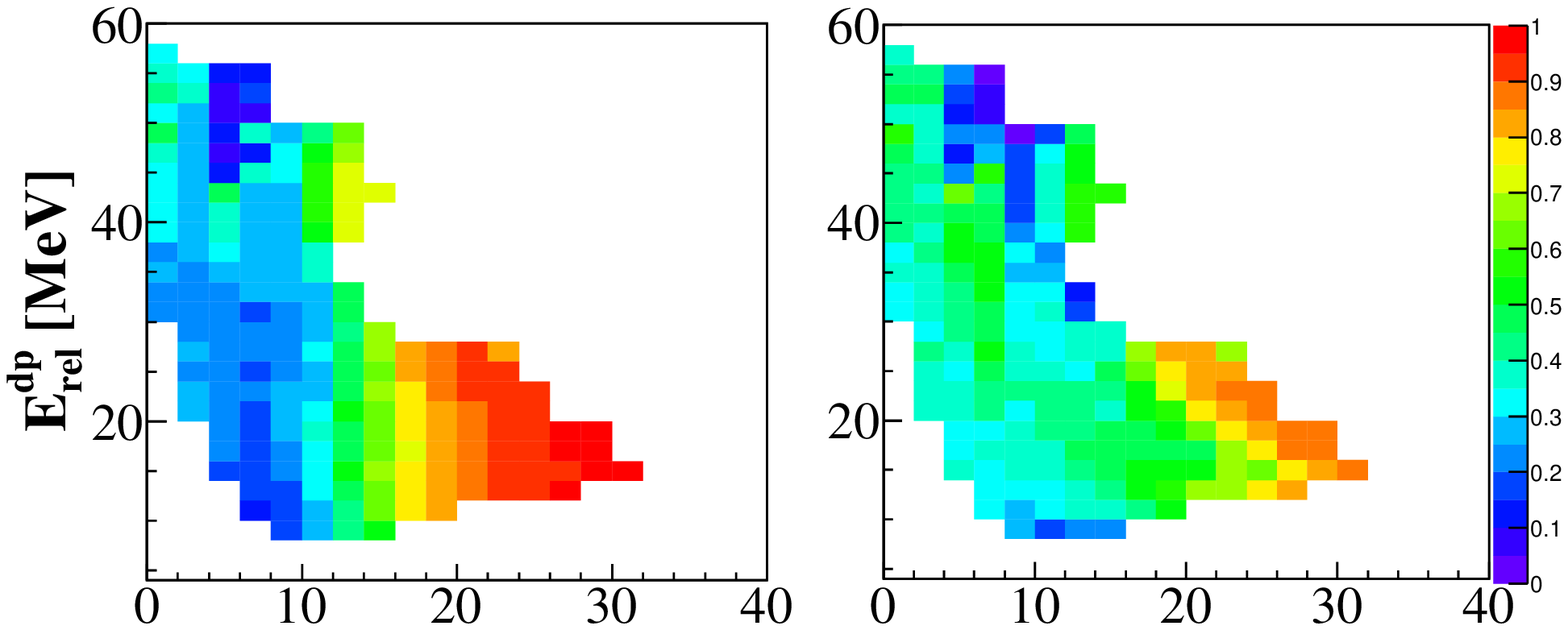}
\includegraphics[width=0.5\textwidth]{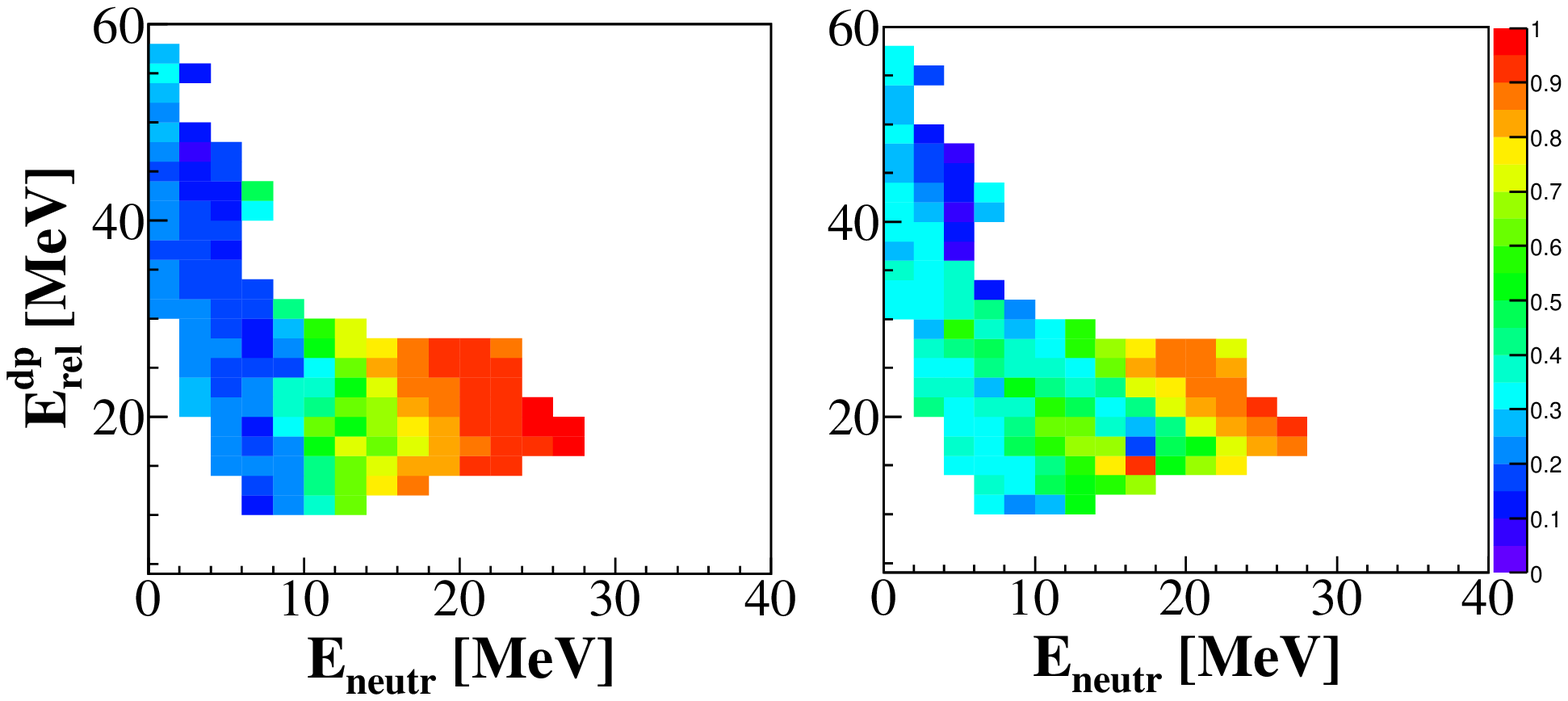}
\caption{Quality of the description in terms of the A-factor given by SSA calculations with various potentials: CDB ({\em upper panel}), CDB+$\Delta$ ({\em middle panel}) 
and  AV18 ({\em bottom panel}). Left and right columns show {\em 1-term} and {\em 4-term} calculations, respectively.}
\label{fig28}
\end{figure}
with exception of combination of $E_{n}<5$ MeV and $E_{d-p}>40$ MeV, as seen in Fig. \ref{fig28}, where also {\em 4-term} predictions
are in agreement with the data. At the lowest $E_{n}<2$ MeV CDB and AV18 perform better than CDB+$\Delta$.\\
\indent The predicted 3NF effects in the QFS regions are small and having the approximate SSA 
calculations no solid conclusions can be drawn about the interaction models.
The SSA provides correct magnitude of the cross-section data, however, it seems 
that the deuteron beam energy of 160 MeV is too low for SSA to provide more accurate results.  

\section{Summary and outlook}
\label{sum}
The differential cross-section distributions for the three-body $^{2}$H({\em d},{\em dp}){\em n} breakup 
reaction have been obtained for 147 proton-deuteron geometries at 160 MeV deuteron beam energy. 
The cross sections have been compared to the calculations based on the single-scattering approximation
for 4N systems at higher energies \cite{Del:16}. The system dynamics 
is modeled with AV18, CD Bonn and CD Bonn+$\Delta$ potentials.
The calculations are still not exact, but they provide correct order of magnitude for the 
cross section near the QFS region. In this region the SSA cross sections are usually higher than the experimental ones, roughly 
by a factor of 2 or 3, with exception of very limited region
in a $E_{d-p}$ vs. $E_{n}$ plane, where the description of the data is satisfactory. \\
\indent From the SSA calculations one could expect much better agreement between experimental and theoretical cross section
at the lowest neutron spectator energies, but this picture may be disrupted due to too low beam energy and 
not negligible final-state interactions.\\ 
\indent The data measured at KVI at the deuteron beam energy of 130 MeV seem not to confirm 
the SSA results for the cross section \cite{Ram:11,Ram:13,Del:16}.
So far the data were only published for one sample configuration ($\theta_{d}=15^{\circ}$, $\theta_{p}=15^{\circ}$, $\varphi_{dp}=180^{\circ}$). 
Since it is highly unlikely that difference in conclusions is caused by 
relatively small difference in beam energy between these two experiments, we presume problems with the data normalization at 130 MeV.\\
\indent The development of models involving 4N systems is ongoing, though exact numerical calculations
for breakup amplitudes are still distant in time given the complexity of the problem.
Current experimental efforts are focused on further development of the 4N database, 
which is very poor especially for the breakup channels.
In particular, an emphasis is placed on investigations of proton-$^{3}\mathrm{He}$ scattering \cite{Cie:16,Sek:18}
since this system is the simplest one where the 3NFs in the channels of total isospin $T=3/2$ can be
studied. Such isospin dependence studies of 3NFs are crucial for
understanding of nuclear systems with larger isospin asymmetry like neutron-rich nuclei,
neutron matter, and neutron stars \cite{Pie:01,Gan:12}.

\begin{acknowledgments}
This work was supported by the Polish National Science Center under
Grants No. 2012/05/E/ST2/02313 (2013-2016) and No. 2016/21/D/ST2/01173 (2017-2020),
and by the European Commission within the Seventh Framework Programme
through IA-ENSAR (Contract No. RII3-CT-2010-262010). A.D. acknowledges the support by 
the Alexander von Humboldt Foundation under Grant No. LTU-1185721-HFST-E.
\end{acknowledgments}

\newpage

\begin{widetext}
\section*{Appendix: breakup cross-section distributions for individual configurations}
\label{app}

\begin{figure*}[!h]
\hspace{-40mm}
\includegraphics[width=1.1\textwidth]{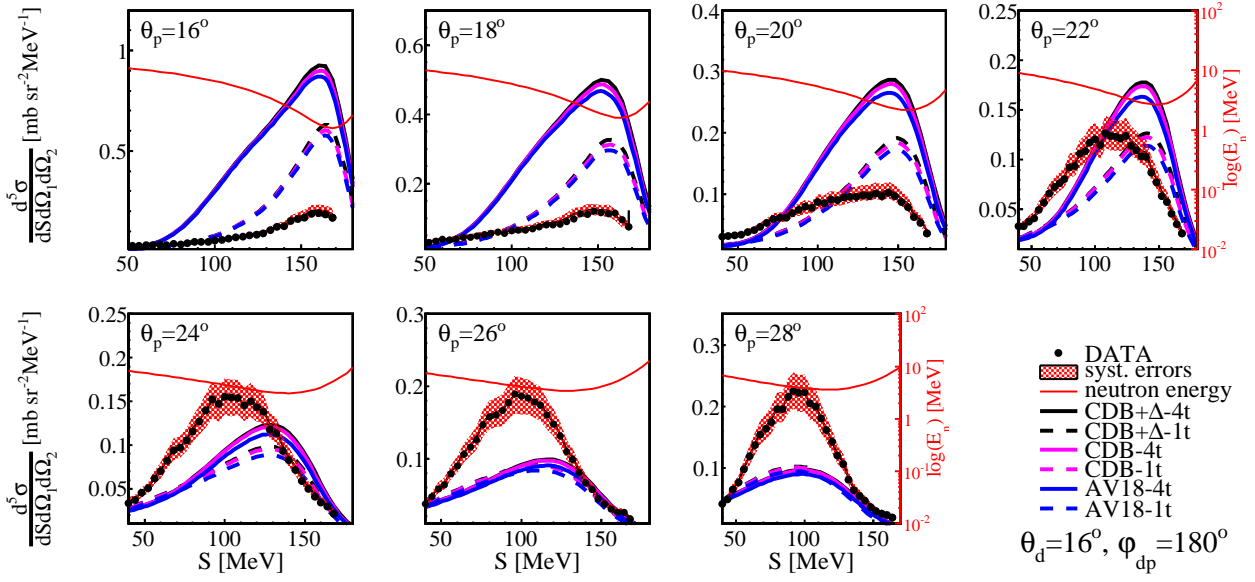}
\caption{Results at $\theta_{d}=16^{\circ}$ and $\varphi_{dp}=180^{\circ}$ for different $\theta_{p}$.
 The dashed lines are for the {\em 1-term} ({\em 1t}) and soli lines for {\em 4-term} ({\em 4t}) calculations based on pure CD Bonn, 
Argonne V18 and CD Bonn+$\Delta$ potentials, as described in the legend.  The red line and the right hand scale present the dependence 
of the spectator neutron energy (E$_{n}$) along {\em S}-axis.}
 \label{fig7}
 \end{figure*}

 
\begin{figure*}
\hspace{-40mm}
\includegraphics[width=1.1\textwidth]{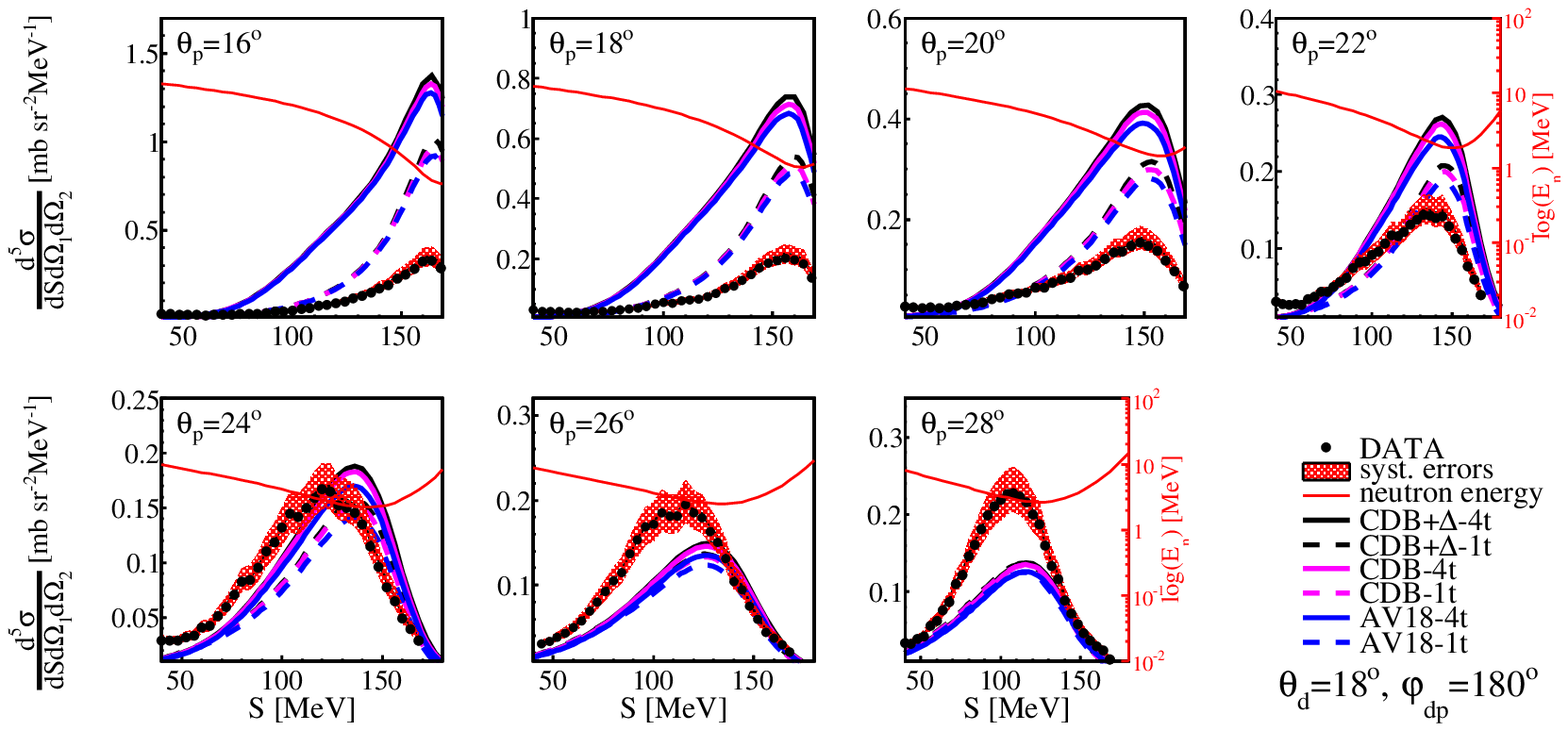}
\caption{The same as Fig. \ref{fig7}, but for $\theta_{d}=18^{\circ}$.}
\label{fig8}
\end{figure*}

\newpage

 \begin{widetext}

 \begin{figure*}
\hspace{-40mm}
\includegraphics[width=0.9\textwidth]{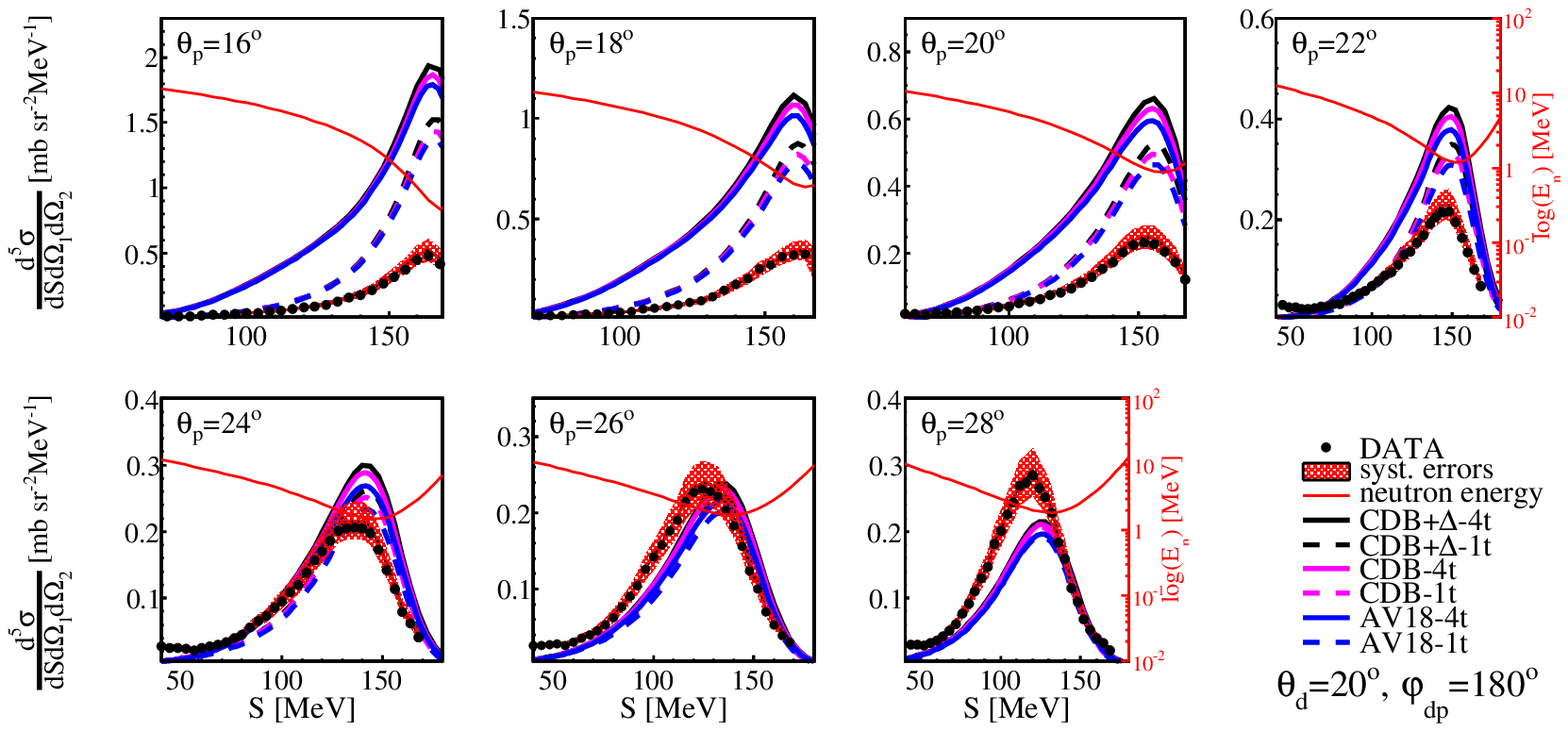}
\caption{The same as Fig. \ref{fig7}, but for $\theta_{d}=20^{\circ}$.}
\label{fig9}
\end{figure*}
 
\begin{figure*}
\hspace{-40mm}
\includegraphics[width=.9\textwidth]{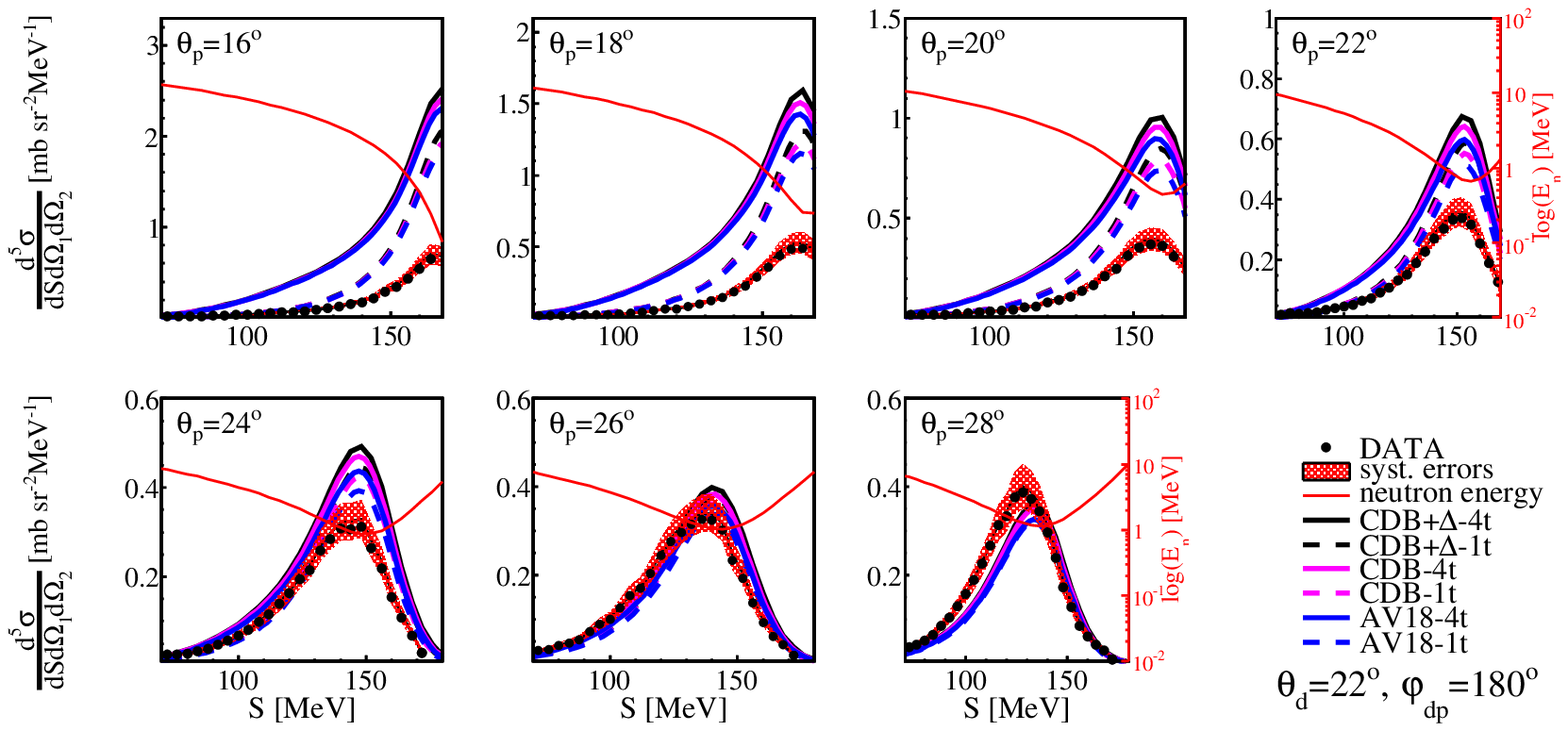}
\caption{The same as Fig. \ref{fig7}, but for $\theta_{d}=22^{\circ}$.}
\label{fig10}
\end{figure*}
 
\begin{figure*}
\hspace{-40mm}
\includegraphics[width=.9\textwidth]{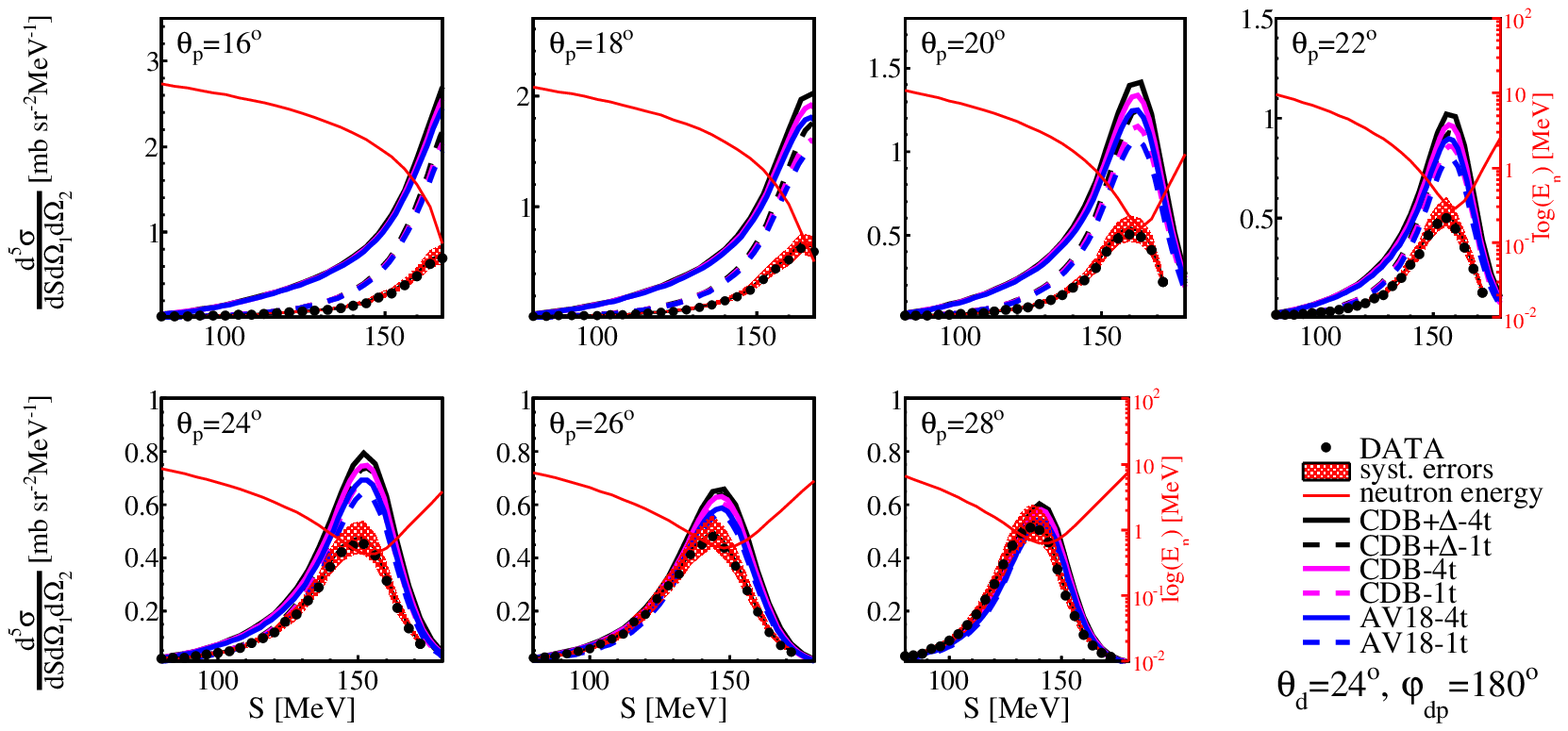}
\caption{The same as Fig. \ref{fig7}, but for $\theta_{d}=24^{\circ}$.}
\label{fig11}
\end{figure*}

\pagebreak
\end{widetext}

 \begin{figure*}
\hspace{-40mm}
\includegraphics[width=0.9\textwidth]{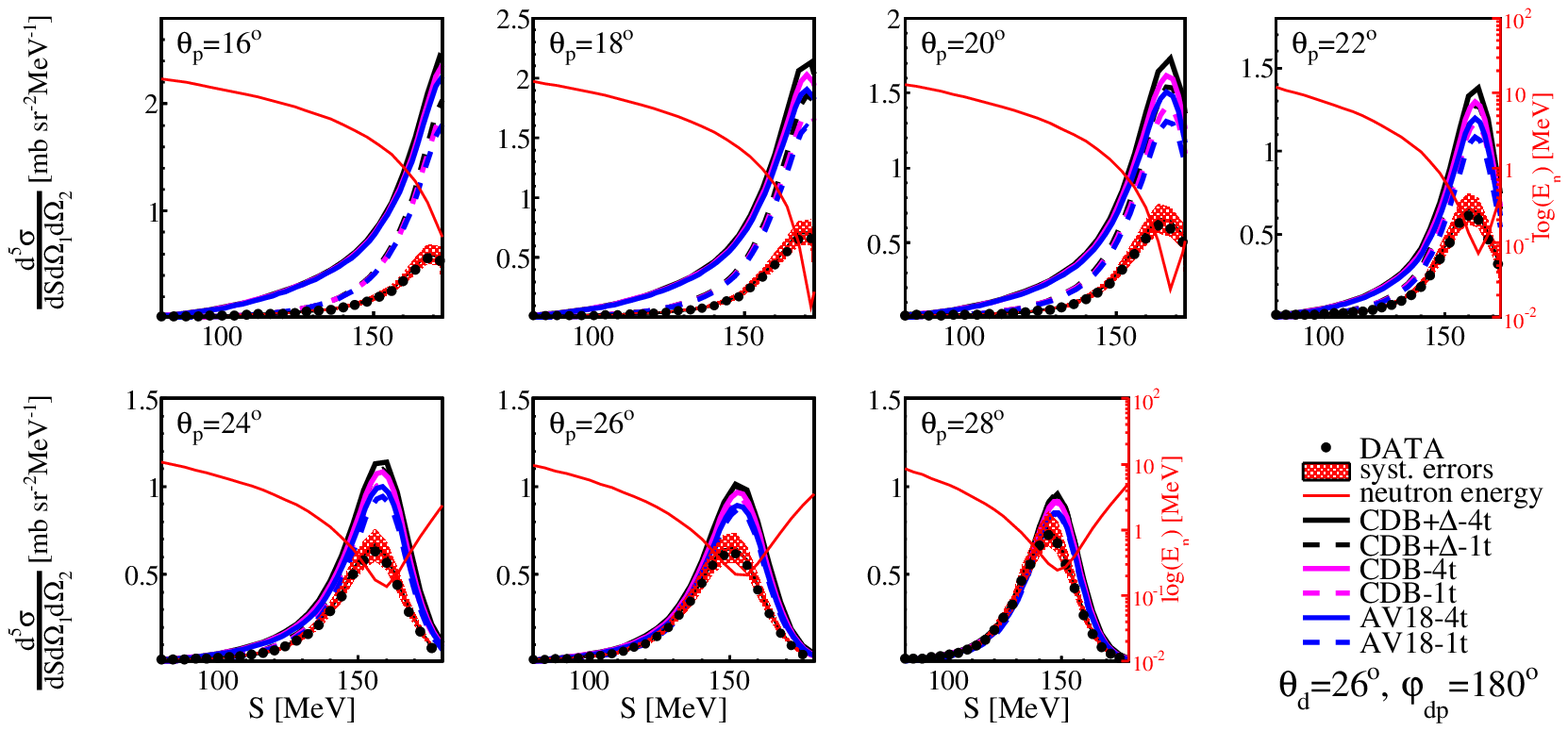}
\caption{The same as Fig. \ref{fig7}, but for $\theta_{d}=26^{\circ}$.}
\label{fig12}
\end{figure*}
  
\begin{figure*}
\hspace{-40mm}
\includegraphics[width=0.9\textwidth]{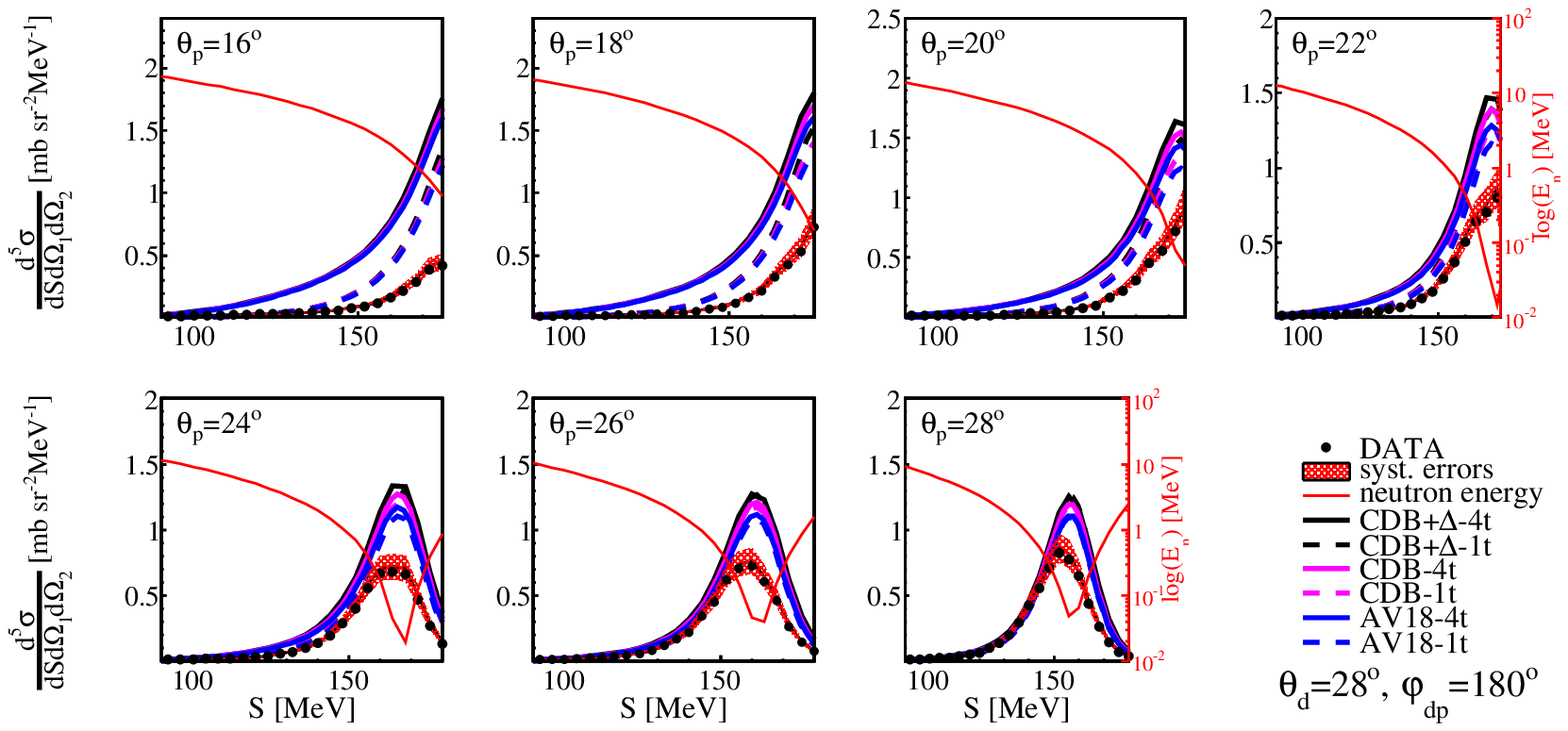}
\caption{The same as Fig. \ref{fig7}, but for $\theta_{d}=28^{\circ}$.}
\label{fig13}
\end{figure*}

\begin{figure*}
\hspace{-40mm}
\includegraphics[width=0.9\textwidth]{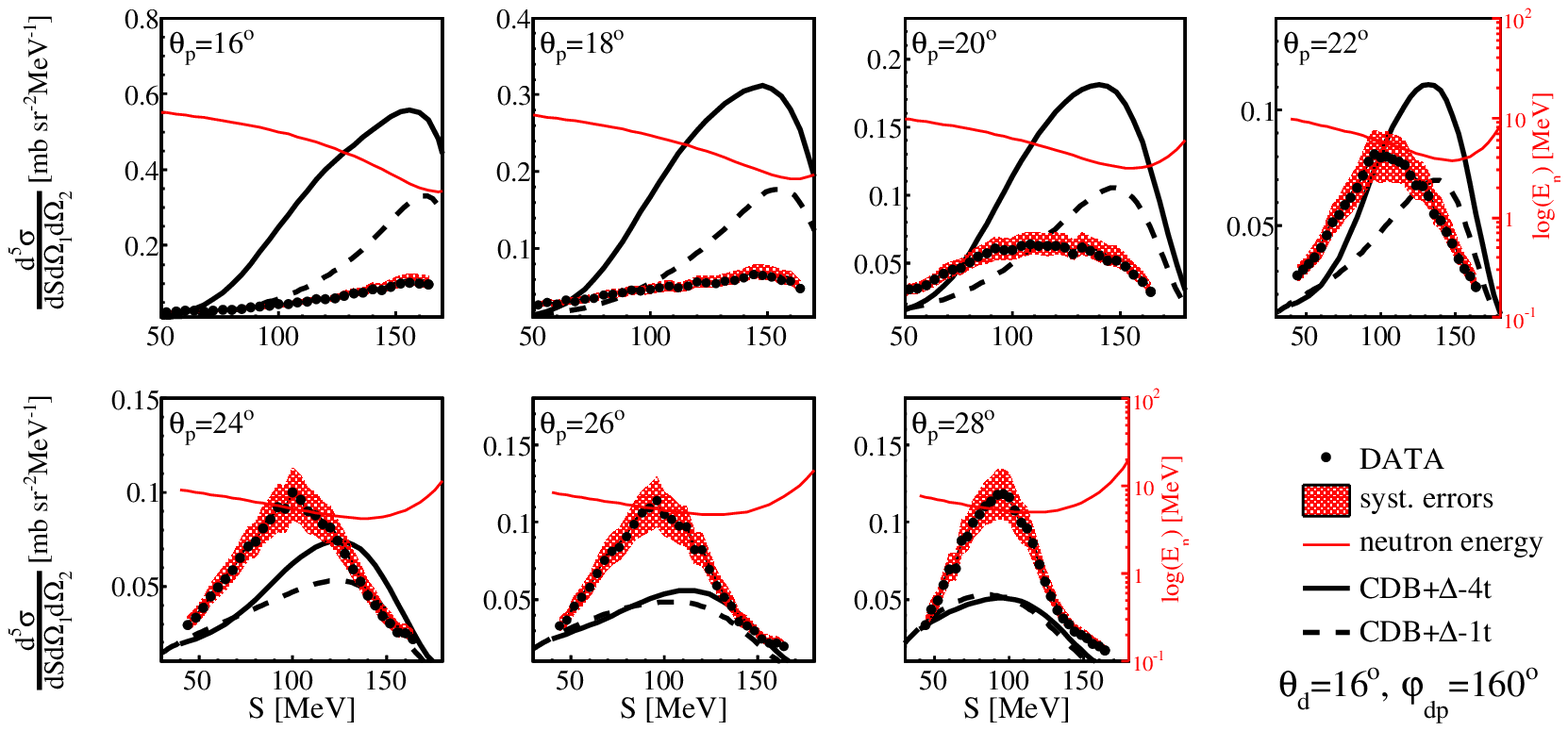}
\caption{Results at $\theta_{d}=16^{\circ}$ and $\varphi_{dp}=160^{\circ}$ for different $\theta_{p}$. The
dashed lines are for the {\em 1-term} ({\em 1t}) and solid lines for  {\em 4-term} ({\em 4t}) calculations 
based on CD Bonn+$\Delta$ potential, as described in the legend. The red line and the right hand scale 
present the dependence of the spectator neutron energy (E$_{n}$) along {\em S}-axis.}
\label{fig14}
\end{figure*}

\newpage

\begin{figure*}
\hspace{-40mm}
\includegraphics[width=0.9\textwidth]{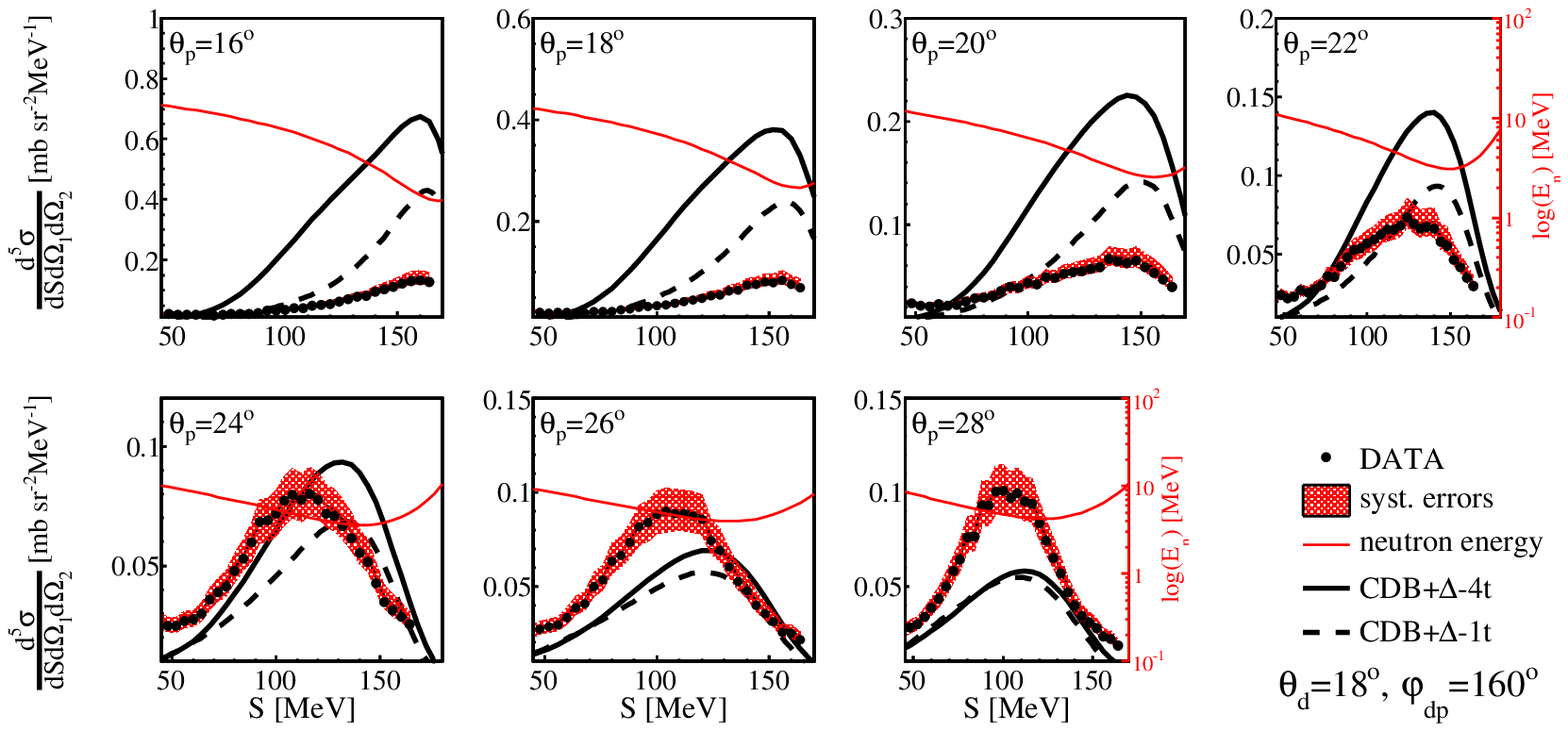}
\caption{The same as Fig. \ref{fig14}, but for $\theta_{d}=18^{\circ}$.}
\label{fig15}
\end{figure*}
 
\begin{figure*}
\hspace{-40mm}
\includegraphics[width=0.9\textwidth]{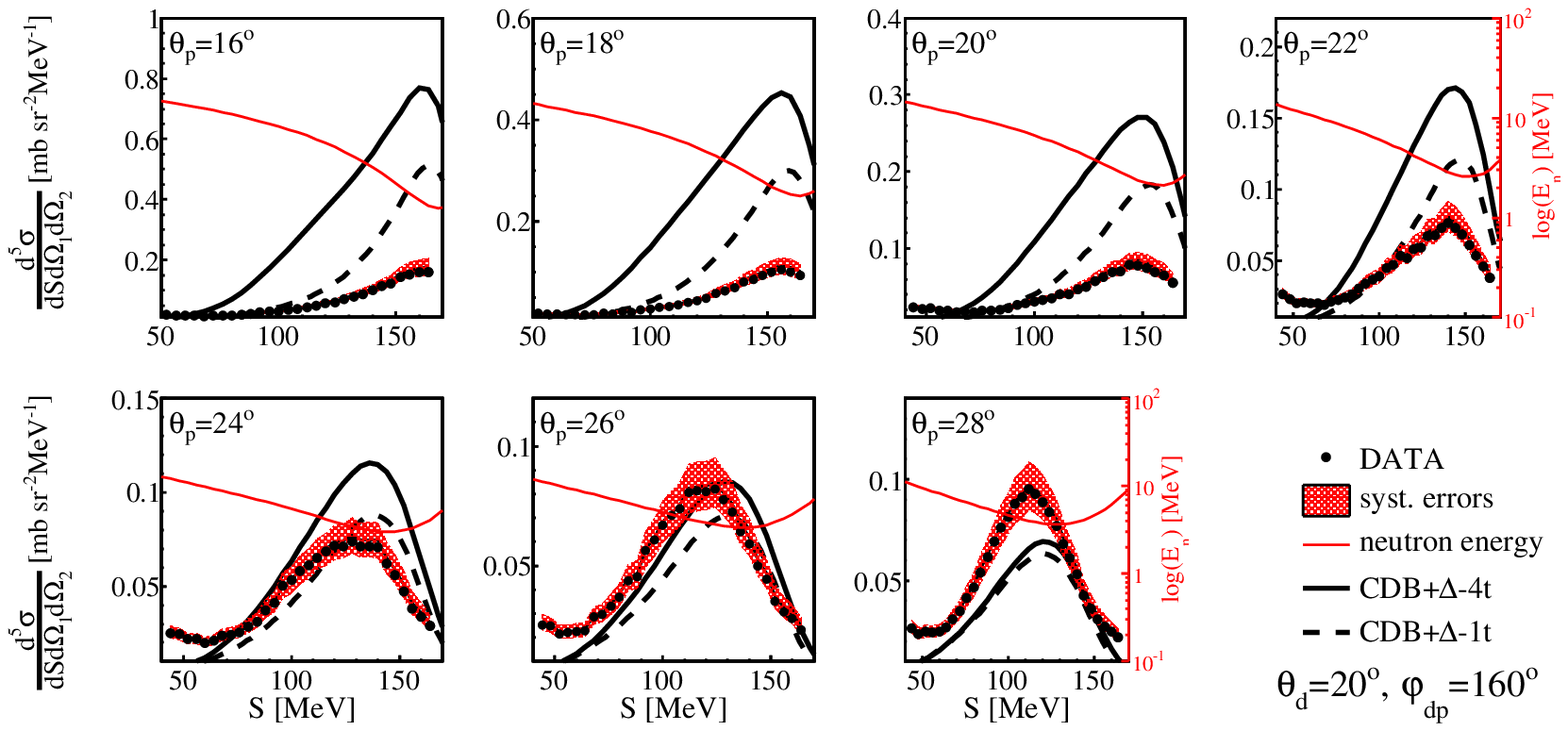}
\caption{The same as Fig. \ref{fig14}, but for $\theta_{d}=20^{\circ}$.}
\label{fig16}
\end{figure*}

\begin{figure*}
\hspace{-40mm}
\includegraphics[width=0.9\textwidth]{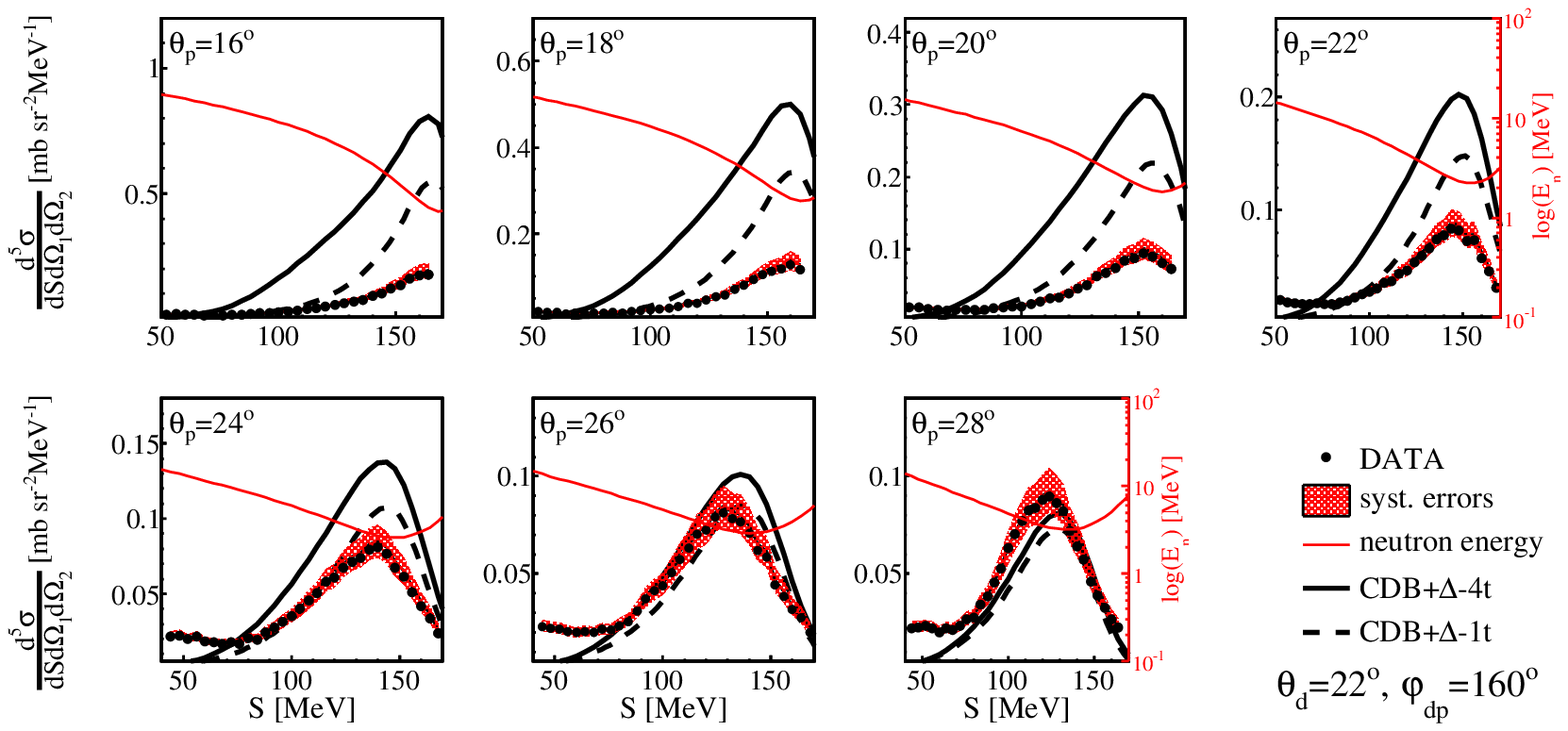}
\caption{The same as Fig. \ref{fig14}, but for $\theta_{d}=22^{\circ}$.}
\label{fig17}
\end{figure*}

\newpage
 
\begin{figure*}
\hspace{-40mm}
\includegraphics[width=0.9\textwidth]{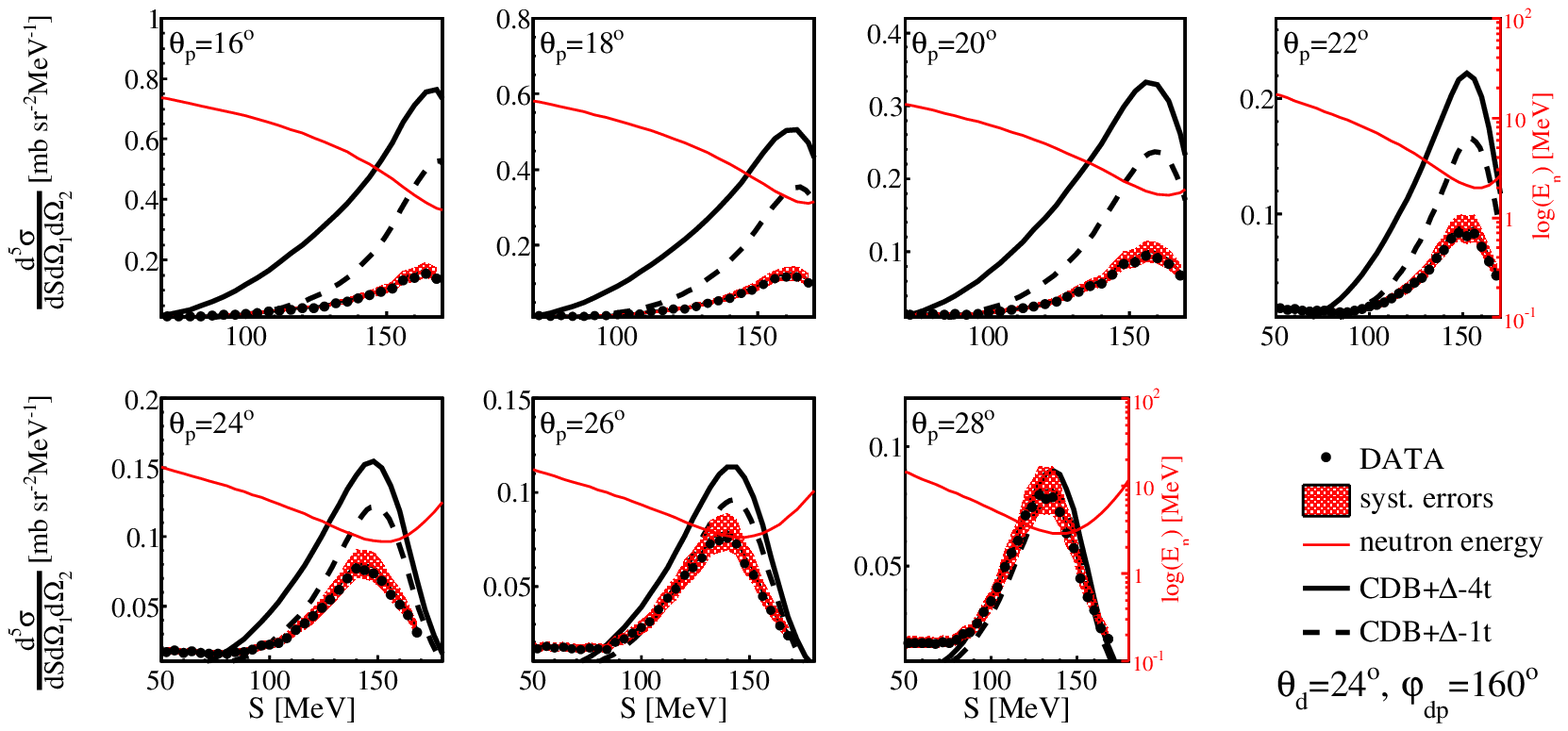}
\caption{The same as Fig. \ref{fig14}, but for $\theta_{d}=24^{\circ}$.}
\label{fig18}
\end{figure*}

\begin{figure*}
\hspace{-40mm}
\includegraphics[width=0.9\textwidth]{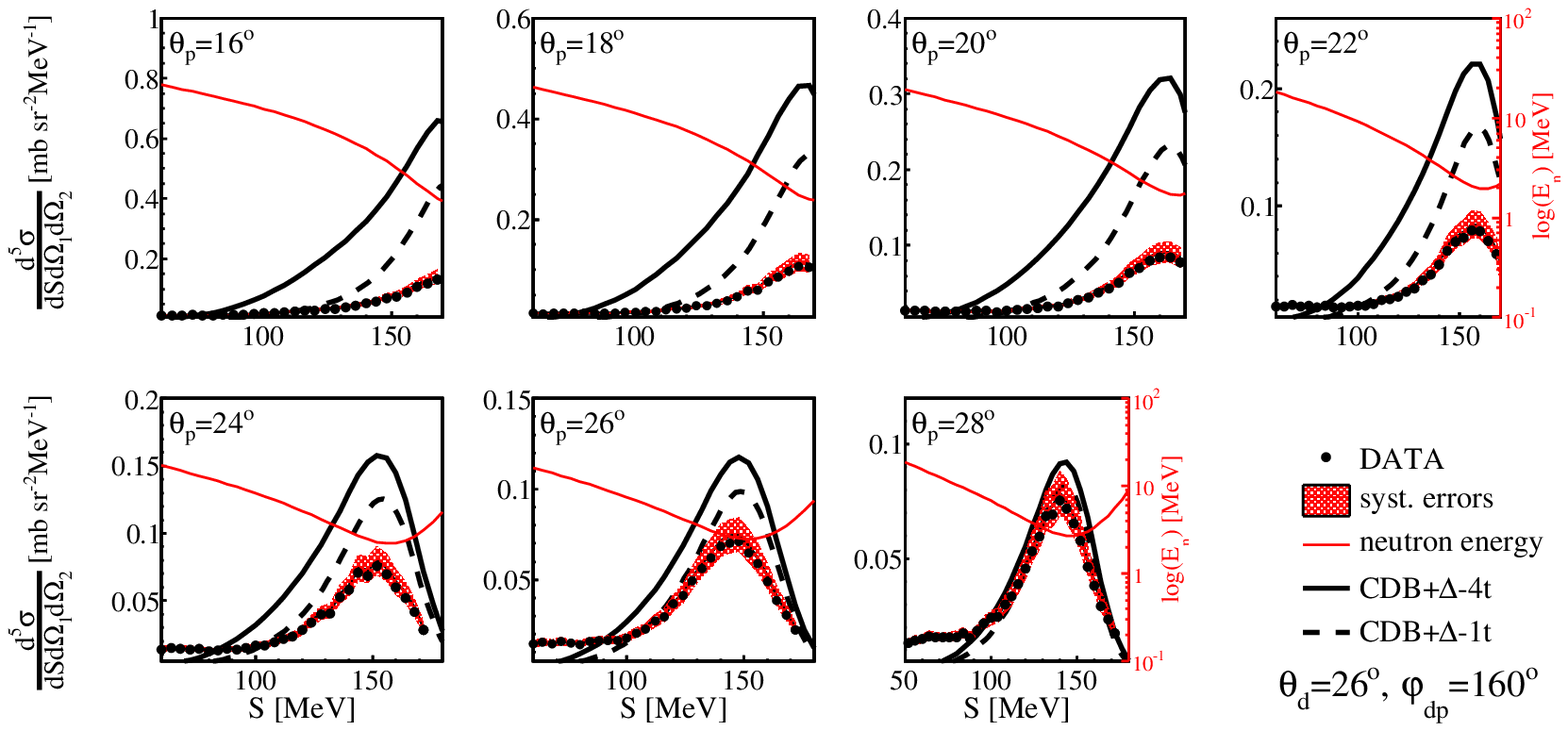}
\caption{The same as Fig. \ref{fig14}, but for $\theta_{d}=26^{\circ}$.}
\label{fig19}
\end{figure*}
 
\begin{figure*}
\hspace{-40mm}
\includegraphics[width=0.9\textwidth]{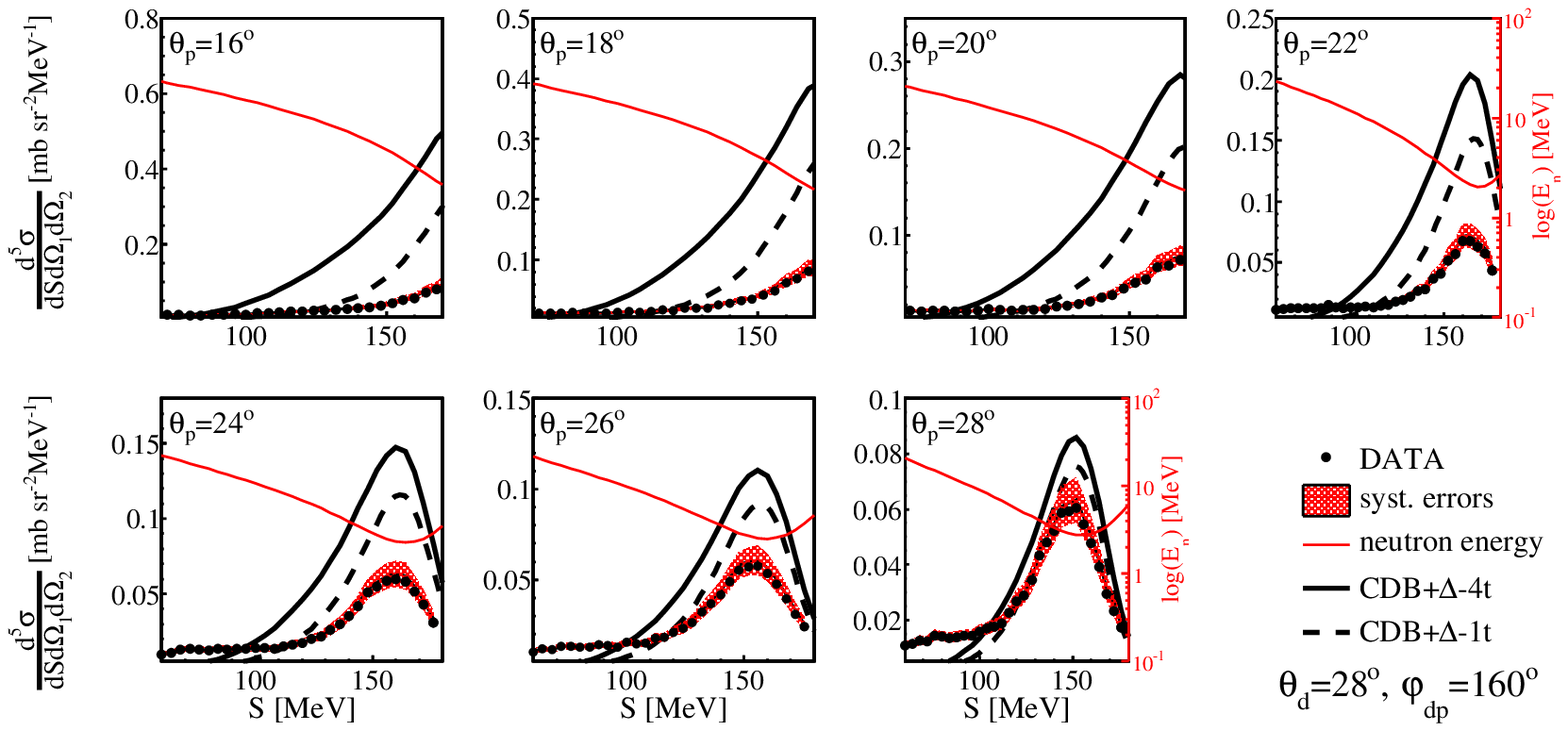}
\caption{The same as Fig. \ref{fig14}, but for $\theta_{d}=28^{\circ}$.}
\label{fig20}
\end{figure*}
 
\newpage
 
\begin{figure*}
\hspace{-40mm}
\includegraphics[width=0.9\textwidth]{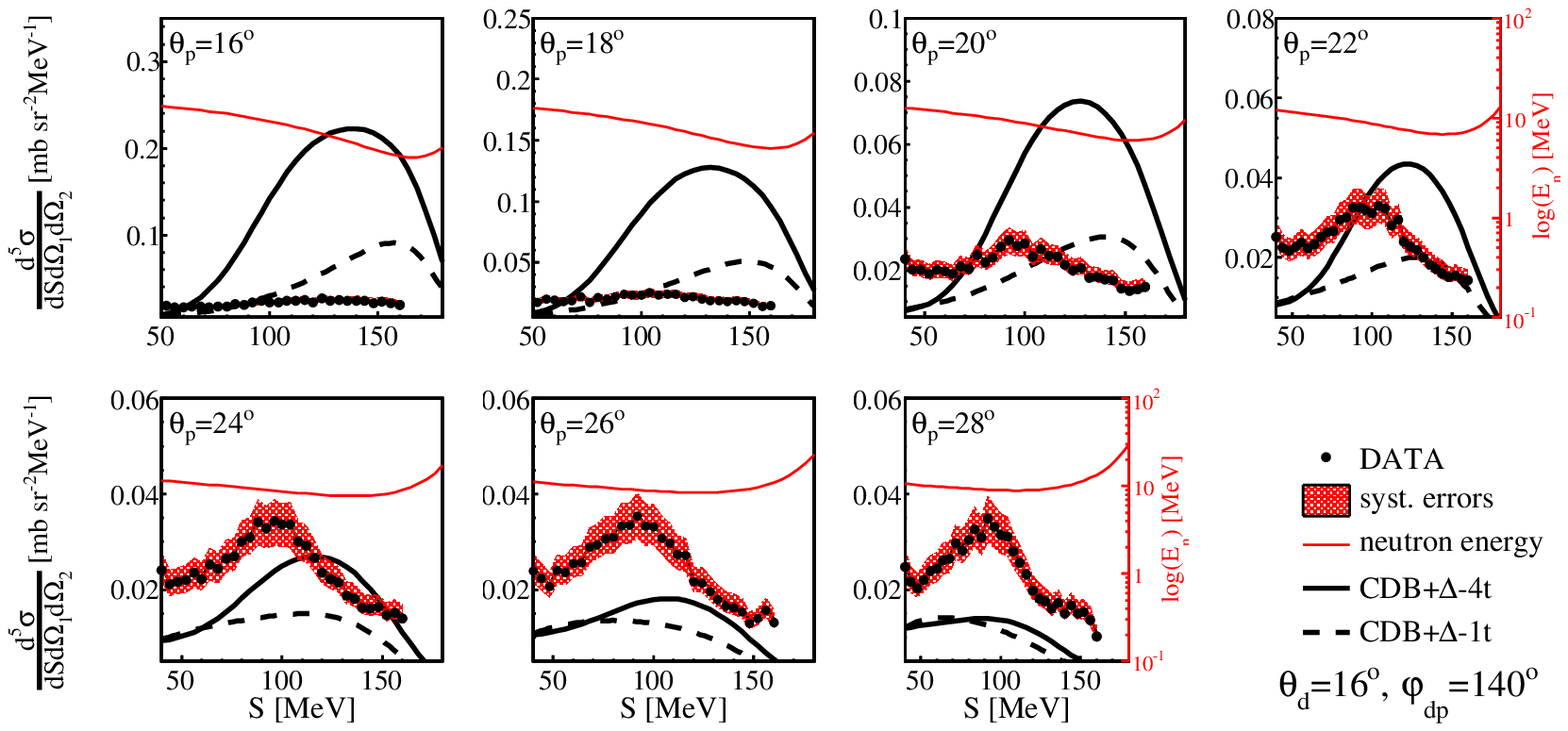}
\caption{Results at $\theta_{d}=16^{\circ}$ and $\varphi_{dp}=140^{\circ}$ for different $\theta_{p}$.
 The dashed lines are for the {\em 1-term} ({\em 1t}) and the solid lines for {\em 4-term} ({\em 4t}) 
calculations based on CD Bonn+$\Delta$ potential, as described in the legend. The red line and the right hand scale 
present the dependence of the spectator neutron energy (E$_{n}$) along {\em S}-axis. }
 \label{fig21}
 \end{figure*}
 
\begin{figure*}
\hspace{-40mm}
\includegraphics[width=0.9\textwidth]{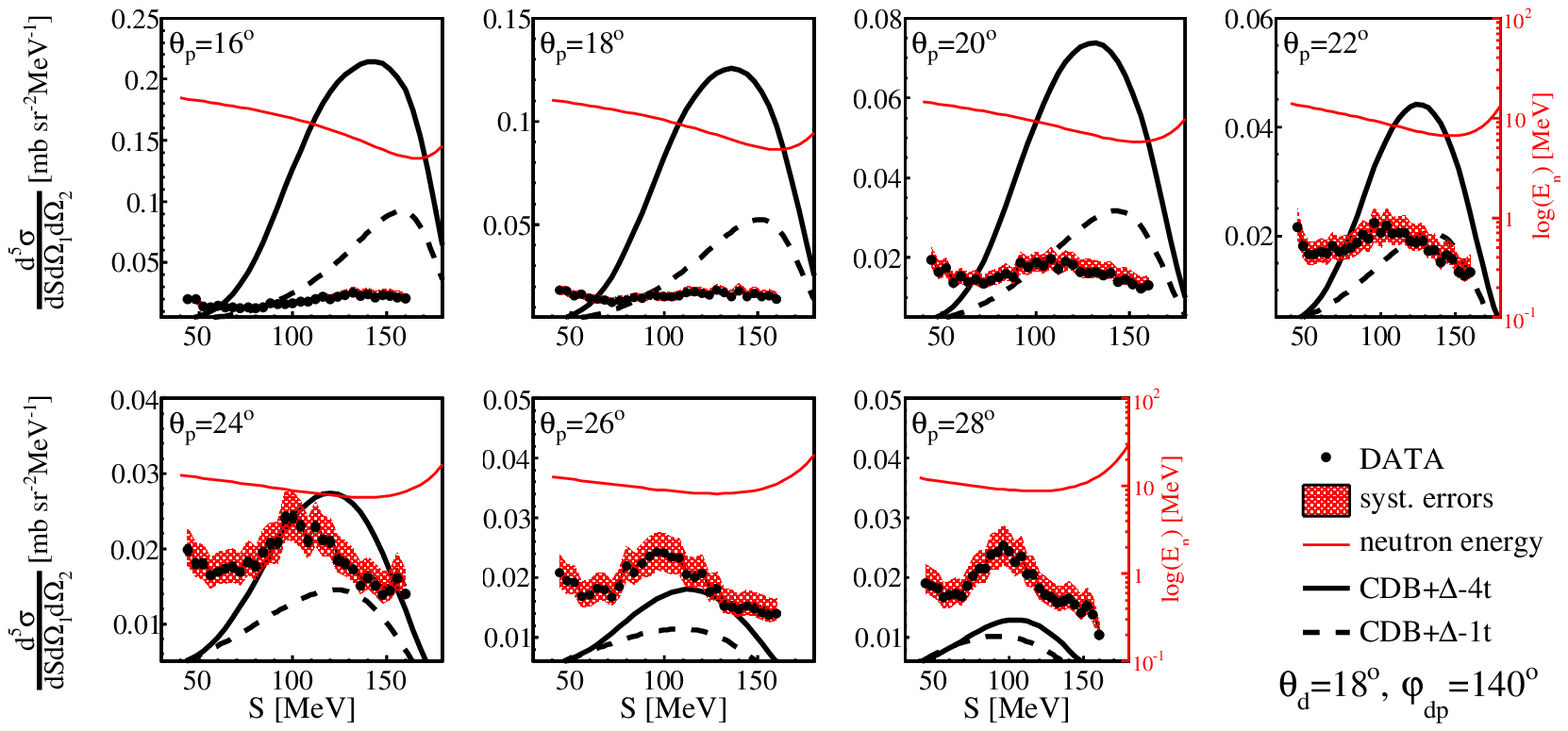}
\caption{The same as Fig. \ref{fig14}, but for $\theta_{d}=18^{\circ}$.}
\label{fig22}
\end{figure*}

\begin{figure*}
\hspace{-40mm}
\includegraphics[width=0.9\textwidth]{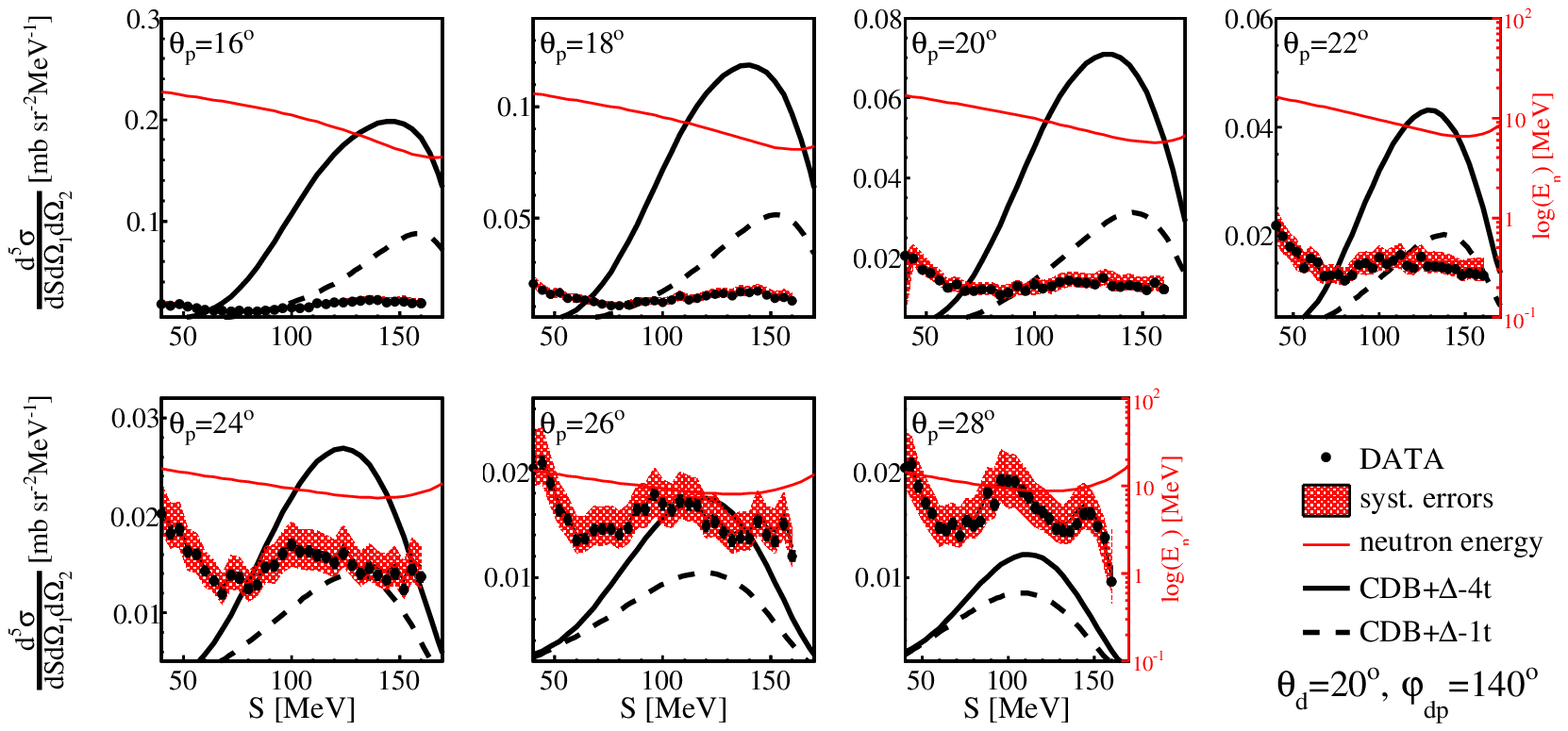}
\caption{The same as Fig. \ref{fig22}, but for $\theta_{d}=20^{\circ}$.}
\label{fig23}
\end{figure*}

\newpage

\begin{figure*}
\hspace{-40mm}
\includegraphics[width=0.9\textwidth]{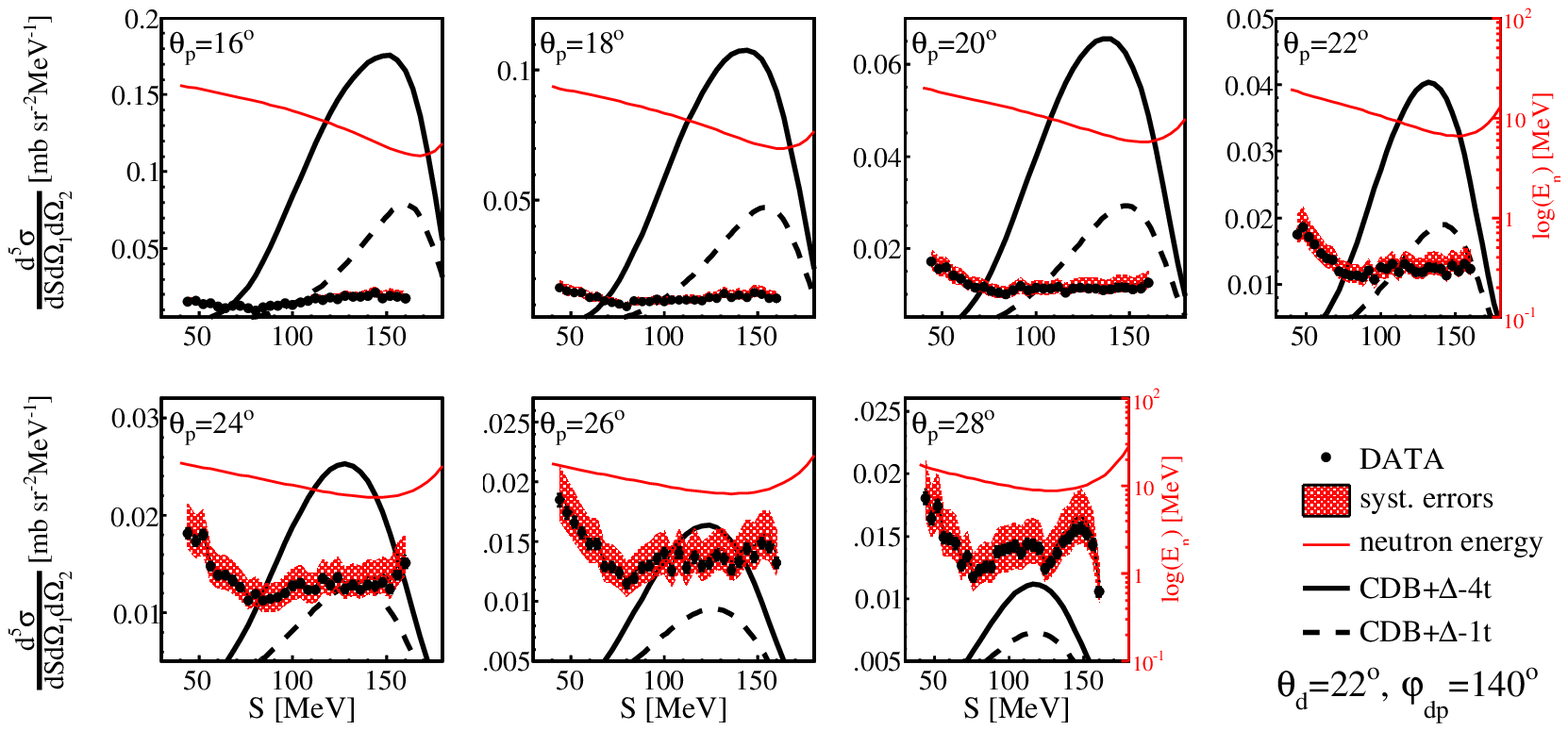}
\caption{The same as Fig. \ref{fig22}, but for $\theta_{d}=22^{\circ}$.}
\label{fig24}
\end{figure*}
 

  
 \end{widetext}


\begin{thebibliography}{99}
\bibitem{Sag:10}K. Sagara et al. Few-Body Syst. {\bf 48}, 59, (2010).
\bibitem{Kal:12}N. Kalantar-Nayestanaki et al., Rep. Prog. Phys. {\bf 75},  016301, (2012).
\bibitem{Kis:13}St. Kistryn and E. Stephan, J. Phys. G: Nucl. Part. Phys. {\bf 40}, 063101, (2013).
\bibitem{Yuk:35}H. Yukawa, Proc. Phys. Math. Soc. Jpn {\bf 17}, 48, (1935).
\bibitem{Wir:95}R. B. Wiringa, V. G. J. Stoks and R. Schiavilla, Phys. Rev. {\bf C51}, 38, (1995).
\bibitem{Mach:01}R. Machleidt, Phys. Rev. {\bf C63}, 024001, (2001).
\bibitem{Stoks:94}V. G. J. Stoks et al., Phys. Rev. {\bf C49}, 2950, (1994).
\bibitem{Faddeev:61}L. D. Faddeev, Sov. Phys. JETP {\bf 12}, 1014, (1961).
\bibitem{Pudli:97}B. S. Pudliner et al., Phys. Rev. {\bf C56}, 1720, (1997).
\bibitem{Coon:01}S. A. Coon, H. K. Han, Few-Body Sys. {\bf 30}, 131, (2001).
\bibitem{Del:03}A. Deltuva, R. Machleidt, P. U. Sauer, Phys. Rev. {\bf C68}, 024005, (2003).
\bibitem{Del:08}A. Deltuva, A. C. Fonseca and P. U. Sauer, Phys. Lett. {\bf B660}, 471, (2008). 
\bibitem{Ent:03}D. R. Entem and R. Machleidt, Phys. Rev. {\bf C68}, 041001(R), (2003).
\bibitem{Ent:17}D. R. Entem, R. Machleidt, Y. Nosyk, Phys. Rev. {\bf C96}, 024004, (2017).
\bibitem{Kis:03}St. Kistryn et al., Phys. Rev. {\bf C68}, 054004, (2003).
\bibitem{Kis:05}St. Kistryn et al., Phys. Rev. {\bf C72}, 044006, (2005).
\bibitem{Kis:06}St. Kistryn et al., Phys. Lett. {\bf B641}, 23, (2006). 
\bibitem{Cie:15}I. Ciepa\l{} et al.,  Few-Body Sys. {\bf 56}, 665, (2015).
\bibitem{Esl:09}M. Eslami-Kalantari et al., Mod. Phys. Lett. {\bf A24}, 839, (2009).
\bibitem{Sekiguchi:09}K. Sekiguchi et al., Phys. Rev. {\bf C79}, 054008, (2009).
\bibitem{Stephan:10}E. Stephan et al., Phys. Rev. {\bf C82}, 014003, (2010).
\bibitem{Ester:13}J. Esterline, W. Tornow, Few-Body Sys. {\bf 54}, 1323, (2013).
\bibitem{Hiy:16}E. Hiyama, R. Lazauskas, J. Carbonell, and M. Kamimura, Phys. Rev. {\bf C93}, 044004, (2016).
\bibitem{Viviani:11}M. Viviani et al., Phys. Rev. {\bf C84}, 054010, (2011).
\bibitem{Kiev:08}A. Kievsky et al., J. Phys. {\bf G35}, 063101, (2008).
\bibitem{Lazauskas:04}R. Lazauskas and J. Carbonell, Phys. Rev. {\bf C70}, 044002, (2004).
\bibitem{Lazauskas:09}R. Lazauskas and J. Carbonell, Phys. Rev. {\bf C79}, 054007, (2009).
\bibitem{Del:14a}A. Deltuva and A. C. Fonseca, Phys. Rev. {\bf C90}, 044002, (2014).
\bibitem{Del:15b}A. Deltuva and A. C. Fonseca, Phys. Lett. {\bf B742}, 285-289, (2015).
\bibitem{Dol:04}P. Doleschall, Phys. Rev. {\bf C69}, 054001, (2004).
\bibitem{Del:13}A. Deltuva and A. C. Fonseca, Phys. Rev. {\bf C87}, 054002, (2013).
\bibitem{Del:16}A. Deltuva and A. C. Fonseca, Phys. Rev. {\bf C93}, 044001, (2016).
\bibitem{Del:17}A. Deltuva and A. C. Fonseca, Phys. Rev. {\bf C95}, 024003, (2017).
\bibitem{Gras:67}E. O. Alt, P. Grassberger and W. Sandhas, Phys. Rev. {\bf C1}, 85, (1967).
\bibitem{Alde:78}C. Alderliesten and A. Djaloeis, Phys. Rev. {\bf C18}, 2001, (1978).
\bibitem{Mich:07}A. Micherdzi\'{n}ska et al., Phys. Rev. {\bf C75}, 54001, (2007).
\bibitem{Bail:09}C. Bailey, Ph. D. thesis, Indiana University, U.S.A., (2009).
\bibitem{Ram:11}A. Ramazani-Moghaddam-Arani et al., Phys. Rev. {\bf C83}, 024002, (2011).
\bibitem{Val:72}V. Valkovi\'{c}  et al., Nucl. Phys. {\bf A183}, 126, (1972).
\bibitem{Cow:74}A. A. Cowley et al., Nucl. Phys. {\bf A220}, 429, (1974).
\bibitem{Yua:77}H. Yuasa, Supplement of the Progress of Theoretical Physics {\bf 61}, (1977).
\bibitem{Klu:78}W. Kluge  et al., Nucl. Phys. {\bf A302}, 93, (1978).
\bibitem{Wiel:81}B.J. Wielinga et al., Nucl. Phys. {\bf A383}, 11, (1981).
\bibitem{Mye:83}L. T. Myers et al., Phys. Rev. {\bf C28}, 29, (1983).
\bibitem{Bizard:80}G. Bizard et al., Phys. Rev. {\bf C22}, 1632, (1980).
\bibitem{Ram:13}A. Ramazani-Moghaddam-Arani et al., Phys. Lett. {\bf B725}, 282, (2013). 
\bibitem{Mar:08}H. Mardanpour, Ph. D. thesis, University of Groningen, (2008).
\bibitem{Stephan:09}E. Stephan et al., Eur. Phys. J. {\bf A42}, 13, (2009).
\bibitem{Stephan:13}E. Stephan et al., Eur. Phys. J. {\bf A49}, 36, (2013).
\bibitem{Kal:98}N. Kalantar-Nayestanaki, J. Mulder, and J. Zijlstra,  Nucl. Instr. Meth. in Phys. Res. {\bf A417}, 215, (1998).
\bibitem{Kal:00}N. Kalantar-Nayestanaki et al., Nucl. Instr. Meth. in Phys. Res. {\bf A444}, 591, (2000).
\bibitem{Cie:18}I. Ciepa\l{} et al., Phys. Rev. {\bf C99}, 014620, (2019).
\bibitem{Khatri:15}G. Khatri, Ph. D. thesis, Jagiellonian University, Krak\'{o}w, (2015).
\bibitem{Parol:14a}W. Parol et al., Acta Phys. Pol. {\bf 45}, 527, (2014).
\bibitem{Kub:18}J. Kubo\'{s} et al., Acta Phys. Pol. {\bf B49}, 451, (2018).
\bibitem{Bir:51}J. B. Birks, Proc. Phys. Soc. {\bf A64}, 874, (1951).
\bibitem{Parol:14}W. Parol et al., EPJ Web of Conferences {\bf 81}, 06007, (2014).
\bibitem{Parol:17}W. Parol et al., Acta Phys. Pol. {\bf B10}, 149, (2017).
\bibitem{Sha:17}S. K. Sharma, B. Kamys, F. Goldenbaum and D. Filges, Eur. Phys. J. {\bf A53}, 150, (2017).
\bibitem{Cie:16}I. Ciepa\l{} et al., Acta Phys. Pol. {\bf B47}, 323, (2016).
\bibitem{Sek:18}K. Sekiguchi et al., JPS Conf. Proc., 012007, (2018).
\bibitem{Pie:01}S. C. Pieper et al., Phys. Rev. {\bf C64}, 014001, (2001).
\bibitem{Gan:12}S. Gandolfi et al., Phys. Rev.  {\bf C85}, 032801, (2012).
 
\end{thebibliography}
\end{document}